\documentclass[trackchanges]{aastex62}

\usepackage{graphicx}
\usepackage{rotating}
\usepackage{hyphenat}
\usepackage{footnote}
\usepackage{url}
\usepackage{color,soul}
\usepackage{lineno}

\usepackage{float}
\usepackage{url}
\usepackage{CJK}

\newcounter{supp}[section]

\submitjournal{PSJ}
\received{2020 August 3}
\revised{2021 March 19}
\accepted{2021 March 23}

\shorttitle{Do rocky planets reflect host\hyp{}star compositions?}

\shortauthors{Schulze et al.}

\graphicspath{{./}{figures/}}

\begin{document}
\begin{CJK*}{UTF8}{gbsn}

\title{On the Probability that a Rocky Planet's Composition Reflects Its Host Star}

\correspondingauthor{Joseph Schulze}
\email{schulze.61@osu.edu}

\author[0000-0003-3570-422X]{J.G. Schulze}
\affiliation{School of Earth Sciences, The Ohio State University, 125 South Oval Mall, Columbus OH, 43210, USA}

\author[0000-0002-4361-8885]{Ji Wang (王吉)}
\affiliation{Department of Astronomy, The Ohio State University, 140 West 18th Avenue, Columbus Ohio, 43210, USA}

\author[0000-0001-7258-1834]{J. A. Johnson}
\affiliation{Department of Astronomy, The Ohio State University, 140 West 18th Avenue, Columbus Ohio, 43210, USA}

\author[0000-0003-0395-9869]{B. S. Gaudi}
\affiliation{Department of Astronomy, The Ohio State University, 140 West 18th Avenue, Columbus Ohio, 43210, USA}

\author[0000-0001-8991-3110]{C.T. Unterborn}
\affiliation{School of Earth and Space Exploration, Arizona State University, 781 Terrace Mall, Tempe, AZ 85287, USA}

\author[0000-0001-5753-2532]{W.R. Panero}
\affiliation{School of Earth Sciences, The Ohio State University, 125 South Oval Mall, Columbus OH, 43210, USA}
\affiliation{Department of Astronomy, The Ohio State University, 140 West 18th Avenue, Columbus Ohio, 43210, USA}

\begin{abstract}
The bulk density of a planet, as measured by mass and radius, is a result of planet structure and composition. Relative proportions of iron core, rocky mantle, and gaseous envelopes are degenerate for a given density. This degeneracy is reduced for rocky planets without significant gaseous envelopes when the structure is assumed to be a differentiated iron core and rocky mantle, in which the core mass fraction (CMF) is a first\hyp{}order description of a planet's bulk composition. A rocky planet's CMF may be derived both from bulk density and by assuming the planet reflects the host star's major rock\hyp{}building elemental abundances (Fe, Mg, and Si). Contrasting CMF measures, therefore, shed light on the outcome diversity of planet formation from processes including mantle stripping, outgassing, and/or late\hyp{}stage volatile delivery. We present a statistically rigorous analysis of the consistency of these two CMF measures accounting for observational uncertainties of planet mass and radius and host\hyp{}star chemical abundances. We find that these two measures are unlikely to be resolvable as statistically different unless the bulk density CMF is at least 40\% greater than or 50\% less than the CMF as inferred from the host star. Applied to 11 probable rocky exoplanets, Kepler\hyp{}107 c has a CMF as inferred from bulk density that is significantly greater than the inferred CMF from its host star (2$\sigma$) and is therefore likely an iron\hyp{}enriched super\hyp{}Mercury. K2\hyp{}229 b, previously described as a super\hyp{}Mercury, however, does not meet the threshold for a super\hyp{}Mercury at a 1\hyp{} or 2\hyp{} $\sigma$ level.

\end{abstract}


\section{Introduction}

\subsection{Major Rock-building Elements in Stars and Planets}
\noindent
Rocky planet composition is degenerate with respect to mass and radius, the primary direct observables of small exoplanets. To break this degeneracy, rocky planets are often assumed to be made of predominantly Fe and MgSiO$_3$ with proportions determined by the relative abundances of the major rock\hyp{}building elements, Fe, Mg, and Si, observed in the host star \citep[e.g.,][]{Dorn15_A&A_bayesian_rocky_interiors,CTU16_ApJ_Scaling_the_Earth, Brugger18_SuperEarth_Interiors}. The primary foundations for these assumptions come from the relationship between the compositions of the solar system rocky planets and the relative solar Fe, Mg, and Si abundances, and the fact that, together with oxygen, these elements make up 95 mol \% of the Earth \citep{McDonough_composition_of_earths_core}.

\vspace{0.25 cm}
\noindent
The Earth's relative bulk composition of major rock\hyp{}building elements reflects that of the Sun \citep{Wang19_ElementalAbund_Vol_Trend}. Upon condensation, the major hosts for Mg and Si are forsterite and enstatite ($\mathrm{Mg_2SiO_4}$ and $\mathrm{MgSiO_3}$) while Fe initially condenses as a metal, each with 50\% condensation temperatures between 1300 and 1350 K \citep{Lodders03}. Further, the most chemically primitive remnants from solar system formation, CI\hyp{}chondrites, have abundances of the refractory and major rock\hyp{}building elements that are within $\pm$ 10$\%$ the relative abundances found in the Sun \citep{Lodders03}. Their Fe/Mg and Si/Mg ratios reflect the solar photospheric ratios to within 2$\%$ and 4$\%$, respectively \citep{Putirka_Rarick_AmM19}.

\vspace{0.25 cm}
\noindent
Mars, like Earth, has a molar Fe/Mg ratio to within $\sim 10-15\%$ of the Sun's abundance \citep{Wanke94_Chem_and_Acc_Hist_Mars, Bertka_Fei_1998, Lodders03, McDonough_composition_of_earths_core, Zharkov_Gudkova_2005, Yoshizaki_McDonough20}. While the Fe/Mg ratio for Venus is poorly constrained, it is consistent with the Earth \citep{Zharkov83_IntStruct_Venus}. Thus, the bulk chemical compositions of Venus, Earth, and Mars appear to be consistent with the hypothesis that these planets initially formed from chondrites, and are thus reflective of the initial relative abundances of the major rock\hyp{}building elements of the solar photosphere.

\vspace{0.25 cm}
\noindent
In contrast, Mercury has an Fe concentration $\sim$200$\%$ \hyp{} 400$\%$ greater than expected relative to silicates \citep[e.g.,][]{Morgan1980}. Therefore, not all rocky planets in the solar system reflect the relative solar abundances of the major rock\hyp{}building elements. The case for Mercury's chemical anomaly suggests an opportunity to study the diversity of the outcomes of planet formation: searching for the chemical anomaly in a large sample of exoplanets.

\vspace{0.25 cm}
\noindent
Starting from a hypothesis that rocky planet compositions mirror their host star's major rock\hyp{}building element abundances, model\hyp{}dependent planet masses and radii can be inferred. For example, iron enrichment relative to magnesium and silicon is invoked to explain higher\hyp{}than\hyp{}expected density \citep[e.g., ][]{Santerne18_K2229b}. Given the hypothesis of compositional mirroring of its host star, a star with a relatively high or low Fe/Mg ratio will form relatively Fe\hyp{}rich (denser) or Fe\hyp{}poor (less dense) planets, respectively. Therefore, planets whose apparent relative iron content is statistically greater than predicted by the host require alternative formation and/or evolutionary mechanisms to explain their compositions. Where the model\hyp{}based masses and radii result in lower densities than predicted from host\hyp{}star abundances, these planets are suggested to have thick surface ice/water layers \citep{CTU18_Nature_Trappist}, be planets enriched in those minerals condensing at the highest temperatures with depleted iron \citep{Dorn_CAI_Planets}, be core\hyp{}free planets \citep{Elkins_Tanton_2008}, or be magma ocean planets \citep{Bower19MagmaOceanPlanets}.

\vspace{0.25 cm}
\noindent In this work, we present a rigorous statistical method to test the null hypothesis, $\mathcal{H}^0$: the measured mass and radius of a given exoplanet is statistically consistent with a model of a barren planet consisting of only an iron core and iron-free silicate mantle in proportions identical to the host star's measured photospheric Fe/Mg and Si/Mg abundance ratios. This approach determines the likelihood that a given planet with well-measured mass and radius satisfies or refutes the oft-invoked assumption that small, dense planets reflect the relative abundances of the major rock-building elements of its host star. In cases where $\mathcal{H}^0$ fails, the planet may either have significant atmospheric layers or non-stellar relative abundances of Fe, Mg, and Si, but our approach makes no attempt to infer the cause.

\vspace{0.25 cm}
\noindent In the case of planets that do not satisfy the null hypothesis, we discuss the range of possible interpretations, including whether such a planet requires a superstellar iron abundance relative to its host or is instead consistent with either a smaller\hyp{}than\hyp{}expected core or an outer volatile layer.

\section{Sample Selection}
\noindent
To test the hypothesis, we identify planets with well\hyp{}constrained mass and radius measurements that are most likely to have rocky surfaces without a significant gas layer. There are over 4000 confirmed planets in the NASA Exoplanet Archive\footnote{\url{https://exoplanetarchive.ipac.caltech.edu/} as of 17 JUN 2020.}, in which 761 planets have both mass and radius measurements. From this sample, we choose planets that are unlikely to retain significant H/He envelopes because of their low surface gravity and the radiation received from their host stars~\citep[e.g.,][]{Jin_Mordasini18_Photoevap}. We use the period\hyp{}dependent radius gap~\citep{VanEylen18} as an upper bound on planet radius. This period\hyp{}dependent radius gap corresponds to R$_p \simeq$ 2.3, 1.9, and 1.5 R$_\oplus$ at orbital periods of 1, 10, and 100 days, respectively. We find 74 planets with measured masses that meet this radius criterion.

\vspace{0.25 cm}
\noindent
We further limit our sample to 28 planets with uncertainties of $\leq 20\%$ and $\leq 10\%$ in planetary mass and radius, respectively. Among them, only 11 planets have host stars with reported Fe, Mg, and Si abundance measurements for their host stars. These 11 planets form the sample for our subsequent analyses, and their properties are summarized in Table \ref{table:SAMPLE}.

\vspace{0.25 cm}
\noindent
Of this sample, the orbital periods range from 0.58 to 6.76 days, radii vary from 1.197 to 1.897 $R_\oplus$, and the observational uncertainties in mass and radius range from 4\% to 19\% and 1.5\% to 6.4\%, respectively. The associated uncertainty in bulk density of these planets ranges from 8\% (55 Cnc e) to 21\% (K2\hyp{}229 b). All of the identified planets are in orbit around FGK stars.

\vspace{0.25 cm}
\noindent

\begin{table}[ht]
   \centering
    
    \footnotesize
    \begin{tabular}{|c|c|c|c|c|c|c|c|}
    \hline

     Planet & $R_p$ ($R_\oplus$) & $M_p$ ($M_\oplus$) & M\hyp{}R Source & $P$ (days) & Fe/Mg & Si/Mg & Spect. Source  \\
     \hline
     K2\hyp{}229 b & $1.197^{+0.045}_{-0.048}$ & $2.49^{+0.42}_{-0.43}$ & \citet{Dai_Homogenous_MRs_HotEarths} & 0.58 & 0.78$\pm$0.19 & 1.1$\pm$0.24 & \citet{Santerne18_K2229b}\\
     HD 219134 c & 1.415$\pm$0.049 & 3.96$\pm$0.34 &  \citet{Ligi19_HD219134} & 6.76 & 0.69$\pm$0.25 & 0.98$\pm$ 0.39 & Hypatia Catalog \\
     Kepler\hyp{}10 b & $1.489^{+0.023}_{-0.021}$  & $3.57^{+0.51}_{-0.53}$ & \citet{Dai_Homogenous_MRs_HotEarths} & 0.84 & 0.62$\pm$0.14 &0.83$\pm$0.16  & \citet{Liu16_kep10_abunds}\\
     HD 219134 b & 1.500$\pm$0.057 & 4.27$\pm$0.34  &  \citet{Ligi19_HD219134} & 3.09 & 0.69$\pm$0.25 & 0.98$\pm$ 0.39 & Hypatia Catalog\\
     Kepler\hyp{}107 c  & 1.597$\pm$0.026 &  9.39$\pm$1.77 & \citet{Bonomo19_Kep107_Nature} &  4.9 & 0.75$\pm$ 0.22 & 0.96$\pm$0.23 & \citet{Bonomo19_Kep107_Nature}\\
     HD 15337 b & $1.699^{+0.062}_{-0.059}$ & $7.20\pm 0.81$  & \citet{HD15337_MR_Dumusque} & 4.76 & 0.69$\pm$0.29 & 0.87$\pm$ 0.20 & Hypatia Catalog\\ 
     K2\hyp{}265 b & $1.71 \pm 0.11$ & $6.54 \pm 0.84$  & \citet{Lam18_k2265b} & 2.37 & 0.84$\pm0.24$ & 0.92$\pm0.24$ & \citet{Lam18_k2265b} \\
     HD 213885 b  & $1.745^{+0.051}_{-0.052}$  & $8.83^{+0.66}_{-0.65}$  & \citet{Espinoza_HD213885} & 1.008 & 0.81$\pm$0.23 & 0.98$\pm$0.31 & \citet{Espinoza_HD213885} \\ 
     WASP\hyp{}47 e & $1.773^{+0.049}_{-0.048}$  & $6.91^{+0.81}_{-0.83}$  & \citet{Dai_Homogenous_MRs_HotEarths} &  0.79 & 0.76$\pm$0.22 & 1.35$\pm$0.36 & \citet{Hellier12_WASP47} \\
     Kepler\hyp{}20 b & $1.868^{+0.066}_{-0.034}$ & $9.70^{+1.41}_{-1.44}$  & \citet{Buchhave16_Kep20b_MR} &  3.70 & 0.71$\pm$0.27 & 0.90$\pm$0.41 & \citet{Kep20_abundances} \\
     55 Cnc e  & $1.897^{+0.044}_{-0.046}$ & $7.74^{+0.37}_{-0.30}$   & \citet{Dai_Homogenous_MRs_HotEarths} & 0.74 & 0.76$\pm$ 0.32 & 0.87$\pm$0.34 & Hypatia Catalog \\

     \hline
    \end{tabular}
    
    \caption{Selected sample of well\hyp{}characterized exoplanets in order of increasing radius. Host star elemental ratios Fe/Mg and Si/Mg are expressed as molar ratios. For each star, we derive molar ratios of Fe/Mg and Si/Mg using the solar abundances from \citet{Lodders09}. For stars with the Hypatia Catalog \citep{Hinkel14_Hypatia} listed as the Spect. Source, we use the listed median values of [Fe/H], [Si/H], and [Mg/H] and calculate the 1$\sigma$ spread of the data around the median values to estimate their uncertainties.}
    
    \label{table:SAMPLE}
\end{table}

\vspace{0.25cm}

\section{Planetary Structure Calculations} \label{sec:Methods1}

\vspace{0.25cm}
\noindent

\subsection{Calculating CMF}

\noindent To first-order, the composition of a rocky planet can be described by the relative amount of iron to silicates \citep{Plotnykov20}. Assuming all Fe is in the core and all silicates reside in the mantle, the composition of a rocky planet can be quantified by its core mass fraction (CMF) given present-day precision in mass and radius \citep{Dorn15_A&A_bayesian_rocky_interiors, CTU16_ApJ_Scaling_the_Earth}.

\vspace{0.25 cm}
\noindent Therefore, we test $\mathcal{H}^0$ through comparison of two independent calculations of the CMF: (1) the fraction core required to explain the average density of the planet, CMF$_\rho$, and (2) the mass fraction of core as predicted by the Mg, Si, and Fe relative abundances of the star, CMF$_{\mathrm{star}}$. We determine that the hypothesis $\mathcal{H}^0$ is refuted when these two measures for CMF differ given the limits of the observational data.

\vspace{0.25 cm}
\noindent 
We use the thermodynamically self\hyp{}consistent \texttt{ExoPlex}\footnote{\url{https://github.com/CaymanUnterborn/ExoPlex}} mass\hyp{}radius software \citep{CTU18_Nature_Trappist} to solve for CMF$_\rho$. ExoPlex solves the five coupled differential equations: the mass within a sphere, hydrostatic equilibrium, adiabatic temperature profile, Gauss's law of gravity in one dimension, and the thermally\hyp{}dependent equation of state. We fix the planetary mass and set a radius convergence criterion of $0.0001 R_\oplus$, more than two orders of magnitude more precise than the planetary radius uncertainties in our sample.

\vspace{0.25 cm}
\noindent
CMF$_\rho$ calculations assume a pure solid Fe core and an oxidized Fe\hyp{}free silicate mantle (MgSiO$_3$) with a solid surface (i.e., it is not a magma ocean planet). For this calculation, we make a simplifying assumption that the mantle has a fixed molar ratio of Si/Mg = 1. Si/Mg ratios between 0.5 and 2 affect the calculation in planet mass by no more than 2\% \citep{CTU16_ApJ_Scaling_the_Earth, Dorn15_A&A_bayesian_rocky_interiors},  less than the observational uncertainties.  We also do not include minor mantle elements (i.e. Ca and Al) in our models as these also do not significantly affect inferred masses. We adopt the iron Vinet equation of state from \citet{Smith18_Fe_EoS} for the core and the equation of state developed in \citet{Stixrude05_MgSiO3_EoS} for the mantle. In the simplified, two\hyp{}layer model of a rocky planet, CMF$_\rho$ is the mass of iron, $M_{\mathrm{Fe}}$, divided by the mass of the planet, $M_p$,

\setcounter{equation}{0}

\begin{equation}
       \mathrm{CMF}_\rho = M_{core}/M_p = M_{\mathrm{Fe}}/M_p. 
    \label{equ:CMFrho} 
\end{equation}

\vspace{0.25 cm}
\noindent
The expected proportion of Fe in a rocky planet's core and Mg and Si making up the mantle as a mixture of oxides, CMF$_{\mathrm{star}}$, can be expressed as

\begin{equation}
  \label{CMF_star}
  \text{CMF}_{\mathrm{star}} = \frac{\big(\frac{\text{Fe}}{\text{Mg}}\big)m_{\text{Fe}}}{{\big(\frac{\text{Fe}}{\text{Mg}}\big)m_{\text{Fe}} + \big(\frac{\text{Si}}{\text{Mg}}\big)m_{\text{SiO}_2}} + m_{\text{MgO}}} ,
\end{equation}

\noindent where compositions (X/Y) are the stellar molar ratio for elements X and Y, and $m_i$ is the molar mass of species $i$. This approach makes a parallel assumption to the calculation of CMF$_\rho$, assuming the core is pure iron and the mantle reflects fully oxidized Mg and Si. A mantle composed of fully oxidized Mg and Si with a metallic Fe core implies an oxygen content controlled by the Si and Mg content of the planet \citep{CTU17oxidation}. 

\vspace{0.25 cm}
\noindent
Atomic diffusion in main-sequence stars can result in surface abundances that are different than the bulk composition, as some elements experience preferential gravitational settling. However, Mg, Si, and Fe are all expected to be affected by about the same amount in FGK stars \citep[e.g.,][]{Liu19} and the ratios of these elements reflect the bulk stellar composition.

\vspace{0.25 cm}
\subsection{The Impact of Observational Uncertainties on CMF Calculations}

\vspace{0.25 cm}
\noindent
The comparison between a planet's composition as inferred from our simple model and its host\hyp{}star's abundances is limited by the observational uncertainties of planetary mass and radius, as well as the uncertainties in host\hyp{}star abundances. We, therefore, quantify the relationship between these observational uncertainties and the proportional impact on planetary structure as described by the relative proportions of rocky mantle and metallic core (Table \ref{dcmfrhotab} and \ref{dcmfstar}).

\vspace{0.25 cm}
\noindent
For each mass and radius of the planets in our sample (Table \ref{table:SAMPLE}), we calculate the 1$\sigma$ uncertainties in CMF$_\rho$ from the errors in their mass and radius measurements using the joint 1$\sigma$ mass\hyp{}radius elliptical distribution. We sample 1000 mass\hyp{}radius pairs along the 1$\sigma$ M\hyp{}R ellipse, from which we derive the 1$\sigma$ mean density uncertainty and its corresponding $\mathrm{CMF}_\rho$ uncertainty. 

\vspace{0.25 cm}
\noindent The comparable uncertainty in CMF$_{\mathrm{star}}$ with respect to Fe/Mg and Si/Mg is found through a propagation of uncertainties in Equation (\ref{CMF_star}),

\begin{equation}
    \sigma_{\text{CMF}_{\mathrm{star}}} = \sqrt{\bigg(\frac{\partial \text{CMF}_{\mathrm{star}}}{\partial\big(\frac{\text{Fe}}{\text{Mg}}\big)}\;\delta\bigg(\frac{\text{Fe}}{\text{Mg}}\bigg)\bigg)^2 + \bigg(\frac{\partial \text{CMF}_{\mathrm{star}}}{\partial\big(\frac{\text{Si}}{\text{Mg}}\big)}\;\delta\bigg(\frac{\text{Si}}{\text{Mg}}\bigg)\bigg)^2}.
\end{equation}

\noindent These uncertainties are independent of planet size, and a weak function of composition (Table \ref{dcmfstar}).

\vspace{0.25 cm}
\section{Hypothesis testing}
\noindent
We quantify the probability that a planet's composition reflects the major rock\hyp{}building element composition of its host star by calculating the amount of overlap between CMF$_\rho$ and CMF$_{\mathrm{star}}$ normalized by the null hypothesis that both distributions have the same mean values values that we fix here to 0.5,

\begin{equation}
    P({\mathcal{H}}^0) = P(\text{CMF}_\rho = \text{CMF}_{\mathrm{star}}) = \frac{\int^{1}_{0} \phi(\text{CMF}_\rho, \sigma_{\text{CMF}_\rho})\phi(\text{CMF}_{\mathrm{star}}, \sigma_{\text{CMF}_{\mathrm{star}}})d\text{CMF}}{\int^{1}_{0} \phi(0.5, \sigma_{\text{CMF}_\rho})\phi(0.5, \sigma_{\text{CMF}_{\mathrm{star}}})d\text{CMF}}
    \label{equ:probability}
\end{equation}

\noindent
where $\phi(\mathrm{CMF_\rho}, \sigma_{\mathrm{CMF_\rho}})$ and $\phi(\mathrm{CMF_{star}}, \sigma_{\mathrm{CMF_{star}}})$ are the probability distributions of $\mathrm{CMF_\rho}$ and $\mathrm{CMF_{star}}$, respectively. For calculations, we assume that all distributions are Gaussian.

\vspace{0.25 cm}
\noindent
This approach incorporates the mutual uncertainties of mass, radius, and host\hyp{}star abundances, in which the probability is proportional to the similarity between modeled CMF$_\rho$ and predicted CMF$_{\mathrm{star}}$. Large uncertainties increase the likelihood that a planet will be indistinguishable from $\mathcal{H}^0$. A graphical interpretation of Eqn. \ref{equ:probability} is illustrated in Fig. \ref{fig:eq3_schematic}.

\vspace{0.25 cm}
\noindent The calculation of CMF$_\rho$, CMF$_{\mathrm{star}}$, their uncertainties, and P$({\mathcal{H}}^0)$ are calculated with the publicly available \texttt{ExoLens}\footnote{\url{https://github.com/schulze61/ExoLens}} code developed for this work. The code uses inputs of planet mass, radius, stellar abundances, and the uncertainty in each of these observables.

\vspace{0.25 cm}
\noindent If $P(\mathrm{CMF_\rho} = \mathrm{CMF_{\mathrm{star}}}) \leq 32\%$ then we assert a planet deviates from what is expected at the $1\sigma$ significance level. Similarly, if $P(\mathrm{CMF_\rho} = \mathrm{CMF_{\mathrm{star}}}) \leq 5\%$ then a planet deviates from its host star at the $2 \sigma$ significance level. We consider planets that deviate from their host stars at the $2\sigma$ level to be statistically \textit{inconsistent} with the null hypothesis.

\vspace{0.25 cm}
\noindent
Our strict use of [0,1] bounds in Equation (\ref{equ:probability}) without renormalization of $\phi(CMF)$ outside this range accounts for the fact that negative $\mathrm{CMF_\rho}$ values require planets with compositions that are inconsistent with any $\mathrm{CMF_{\mathrm{star}}}$ value (as defined by Equation (\ref{CMF_star})), and thus are inconsistent with our null hypothesis. If we were to renormalize the probability distribution for $\mathrm{CMF_\rho}$ to be equal to unity over [0,1], we would be enforcing that the planet to be rocky by our definition, and thus would not be testing $\mathcal{H}^0$ as defined. Rather, we would be testing the conditional probability that, if the planet is rocky ($\mathrm{CMF_\rho} \ge 0$), that its inferred composition is consistent with that of the host star, i.e., that the probability distribution of $\mathrm{CMF_\rho}$ normalized to unity over [0,1] is consistent with the probability distribution of $\mathrm{CMF_{\mathrm{star}}}$ at the 2$\sigma$ level. While testing this alternative hypothesis is a reasonable approach, it is more restrictive than our approach, as we do not require that the planet is rocky in the sense that its mass and radius can be fit by a model consisting purely of an iron core and iron-free silicate mantle.

\section{Results}
\noindent 
The average molar Fe/Mg for FGK\hyp{}type stars is $0.7\pm0.18$ \citep{unterborn_panero19, Hinkel14_Hypatia} corresponding to CMF$_{\mathrm{star}} = 0.28^{+0.05}_{-0.06}$ consistent with recent results from \citet{Plotnykov20}. For this sample set, we find a relatively narrow range of CMF$_{\mathrm{star}}$ from 0.26 to 0.33, with uncertainties between 0.06 and 0.10. We find a significantly wider range of CMF$_\rho$ from 0.004\hyp{}0.70 and uncertainties from 0.10\hyp{}0.24.

\begin{figure}[p]
    \centering
    \begin{tabular}{ccc}
        \includegraphics[width=.3\textwidth]{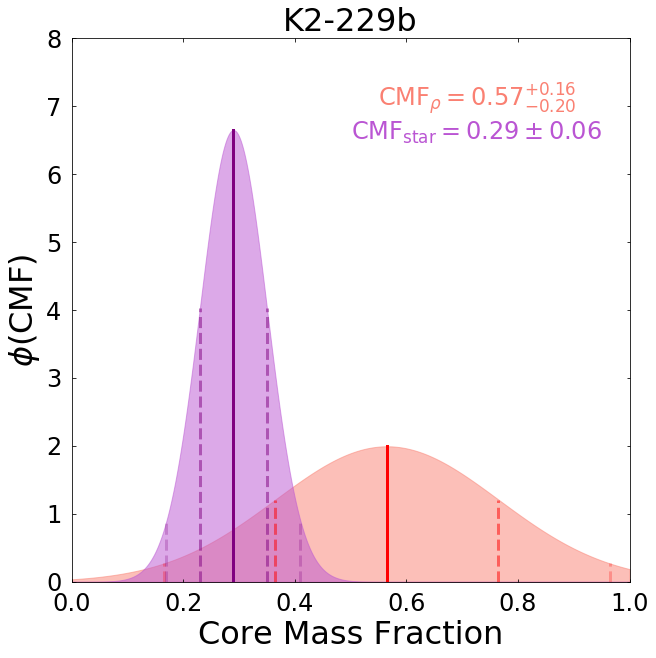} &
        \includegraphics[width=.3\textwidth]{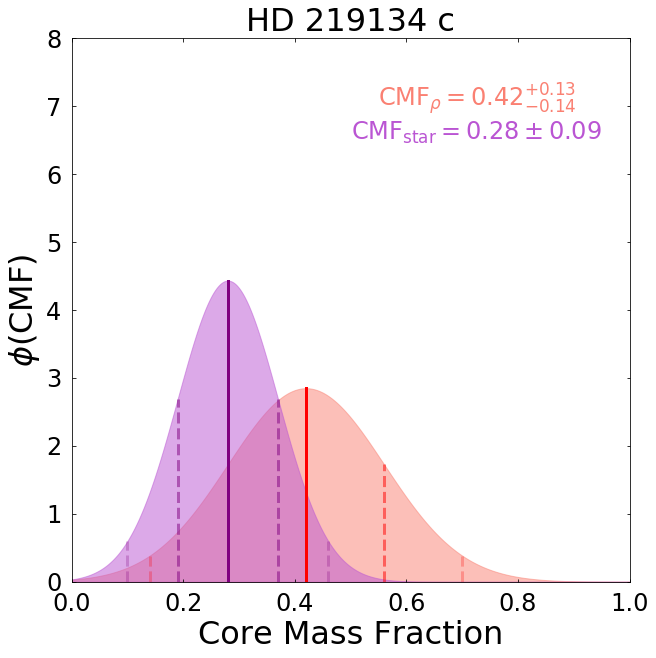} &
        \includegraphics[width=.3\textwidth]{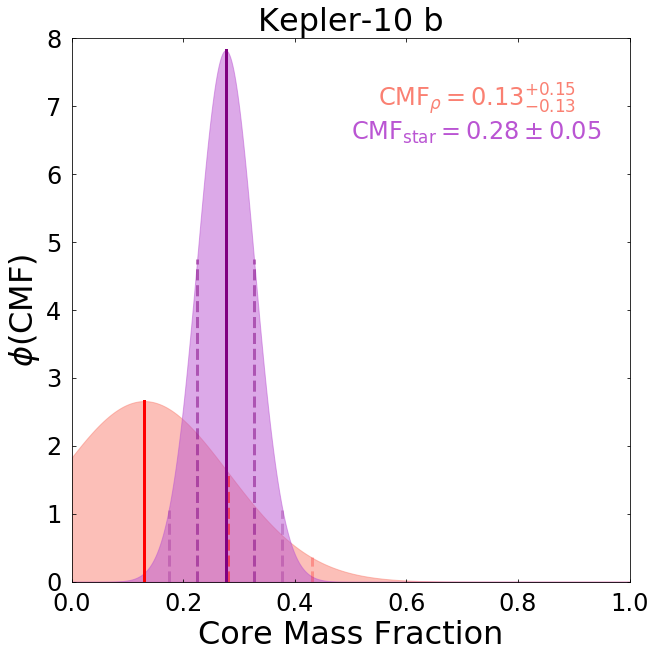} \\
        \includegraphics[width=.3\textwidth]{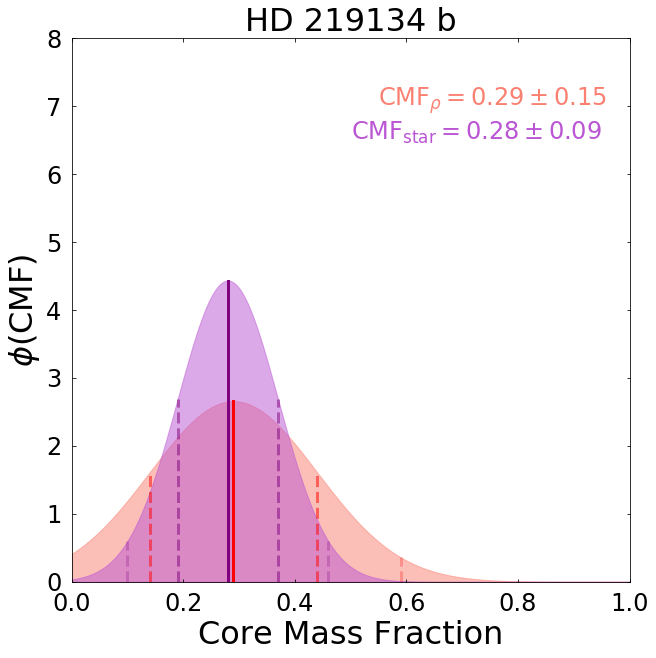} &
        \fcolorbox{cyan}{white}{\includegraphics[width=.3\textwidth]{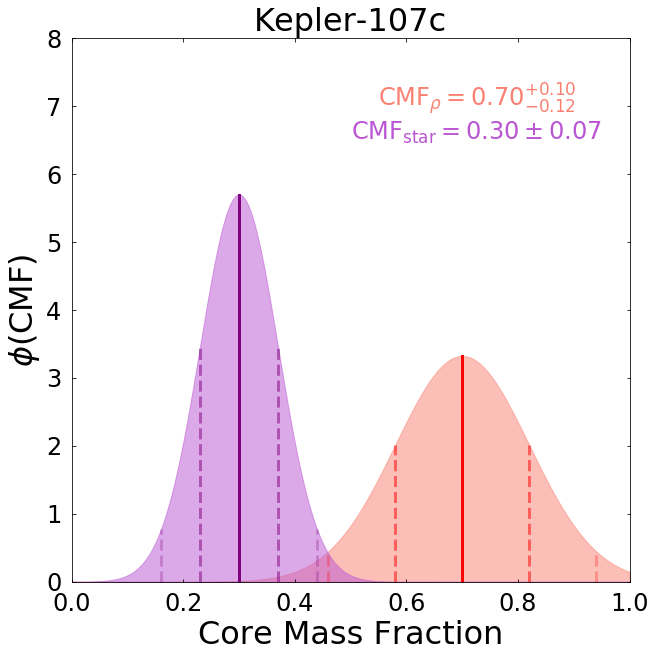}} &
        \includegraphics[width=.3\textwidth]{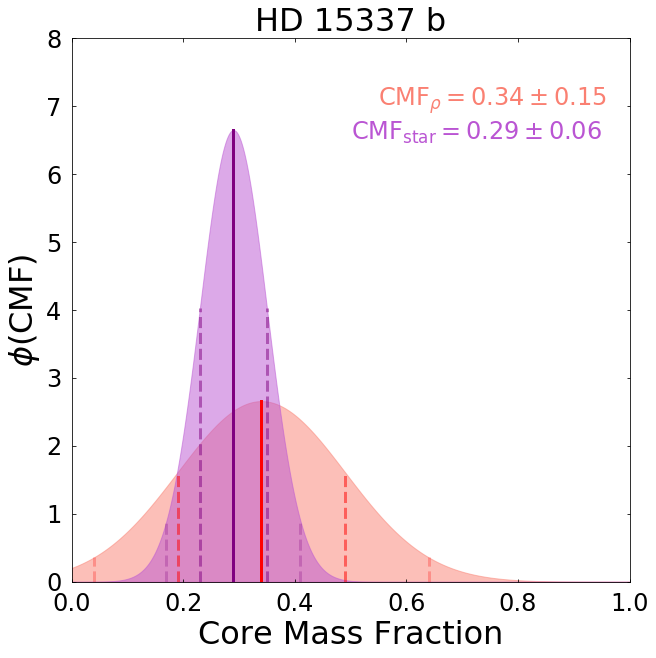} \\
        \includegraphics[width=.3\textwidth]{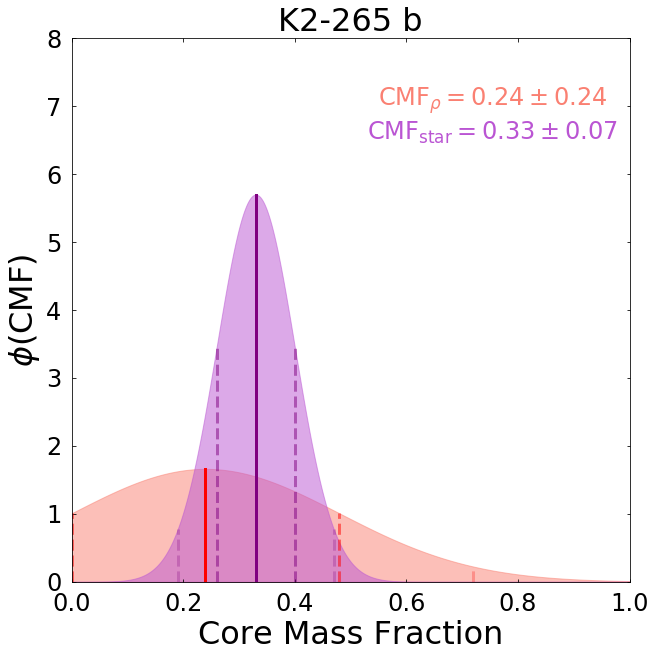} &
        \includegraphics[width=.3\textwidth]{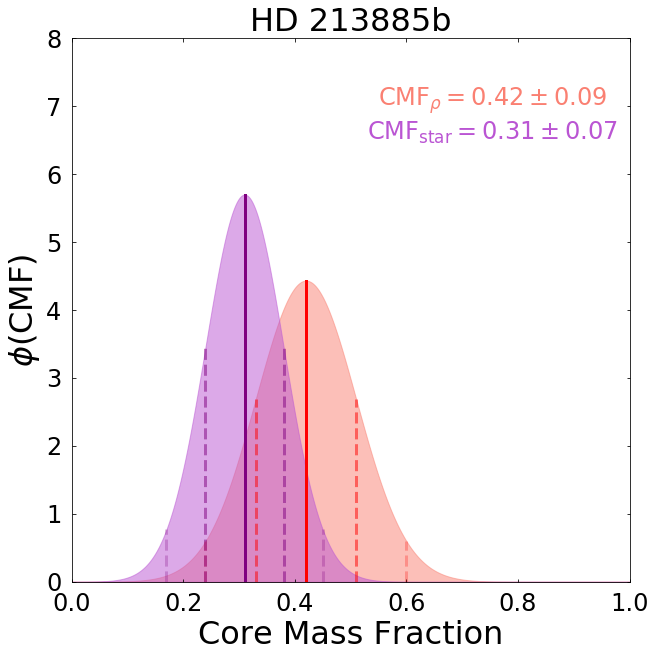} &
        \includegraphics[width=.3\textwidth]{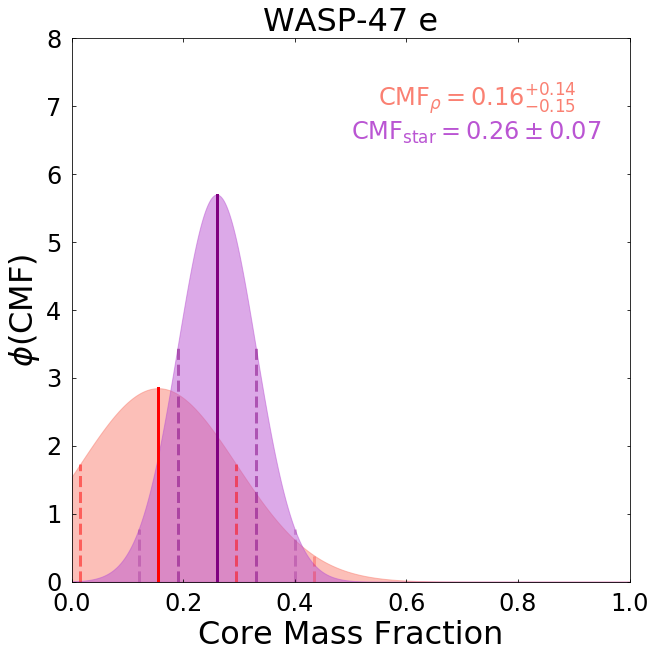} \\
        \includegraphics[width=.3\textwidth]{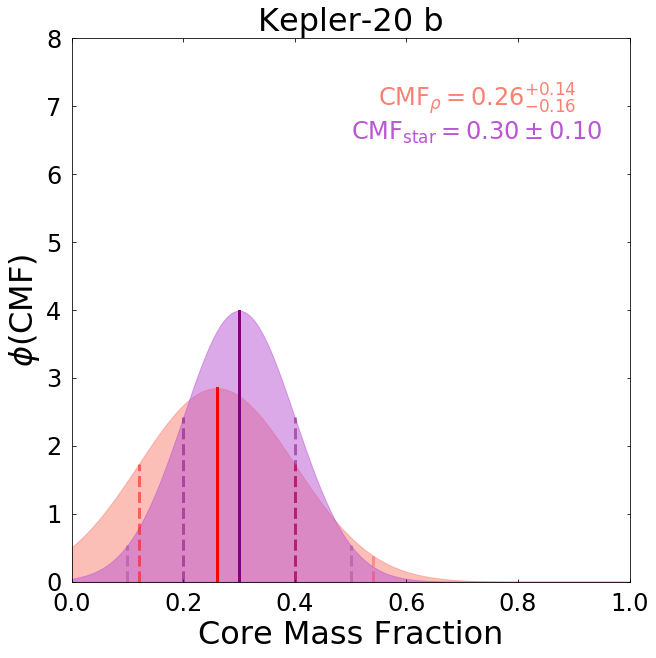} &
        \includegraphics[width=.3\textwidth]{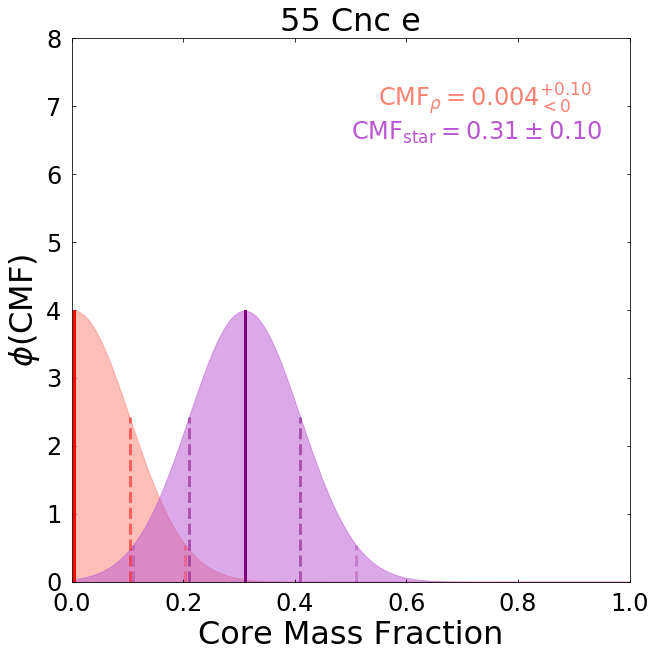} \\     
        
        \end{tabular}

    \caption{Probability density functions ($\phi$) as a function of core mass fraction given 1$\sigma$ distributions for CMF$_\rho$ (red) and CMF$_{\mathrm{star}}$ (purple). All CMF calculations assume that the cores are pure iron and silicate mantles are iron\hyp{}free, such that CMFs with these assumptions are equivalent to iron mass fraction. Figures are as listed in Table \ref{table:SAMPLE}, in order of increasing radius.} 
    
    \label{figure:PDFs}
\vspace{-20pt}
\end{figure}

\vspace{0.25 cm}
\noindent Despite the large range of CMF$_\rho$, the mutual uncertainties of pairs of CMF values are such that all planets but Kepler\hyp{}107c are consistent with the null hypothesis at the $2\sigma$ level (Table \ref{table:RESULTS}; Figure \ref{figure:PDFs}). Therefore, within the limits of the observational measurements, 91\% of planets meeting the selection criteria for this study are not distinguishable from their stellar host in major rock\hyp{}building element composition. The one exception, Kepler\hyp{}107c, has a 1\% likelihood of satisfying $\mathcal{H}^0$, with a CMF$_{\mathrm{star}}$ $<$ CMF$_\rho$ (Fig. \ref{figure:PDFs}) implying greater\hyp{}than\hyp{}expected density relative to that predicted by its host star's major rock\hyp{}building element composition. We, therefore, classify this planet as a super\hyp{}Mercury (SM). The rest of the planets are Indistinguishable from the Host Star (IHS). 

\vspace{0.25 cm}
\noindent
At the $1\sigma$ significance level, the set of those planets inconsistent with the null hypothesis ($P(\text{H}^0) \leq  32 \%$) increases by one (Table \ref{table:RESULTS}). 55 Cnc e has a 9\% probability of being consistent with the null hypothesis, in which CMF$_{\mathrm{star}}$ $>$ CMF$_\rho$. This suggests 55 Cnc e has a lower\hyp{}than\hyp{}expected density relative to its host star's major rock\hyp{}building element abundances given the a priori assumption that the planet is rocky. These results are consistent with previous results that the planet has a potential lower\hyp{}than\hyp{}expected density  \citep[e.g.,][]{Dai_Homogenous_MRs_HotEarths, Dorn_CAI_Planets, Bourrier18_55Cnce, Crida18a_55Cnce, AngeloHu_55Cnce, Demory16_55Cnc, Demory16b_55Cnce}. The failure of the null hypothesis does not assess the cause of the inferred low density as being a result of a primary or secondary gaseous envelope or of a chemical deviation from the major rock\hyp{}building element abundances of the host star.

\begin{table}[ht]
    \centering
    
    \begin{tabular}{|c|c|c|c|c|c|}
    
        \hline
        Planet & CMF$_\rho$ & CMF$_{\mathrm{star}}$ & $P(\mathcal{H}^0)$ (\%) & 1$\sigma$ Class & 2$\sigma$ Class \\
     
        \hline
        
        K2\hyp{}229 b  & $0.565^{+0.16}_{-0.20}$ & $0.29 \pm 0.06$ & 42  & IHS & IHS  \\
        HD 219134 c & $0.42^{+0.13}_{-0.14}$ & $0.28\pm 0.09$& 70 & IHS & IHS   \\
        Kepler\hyp{}10 b & $0.13^{+0.15}_{-0.13}$ & $0.28\pm 0.05$ &  65  & IHS & IHS \\
        HD 219134 b & $0.29\pm0.15$ & $0.28\pm 0.09$ & 100 & IHS & IHS  \\
        Kepler\hyp{}107 c & $0.70^{+0.10}_{-0.12}$ & $0.30\pm0.07$ & 1 & SM & SM\\
        HD 15337 b & $0.34 \pm 0.15$ & $0.29\pm0.07$ & 96 & IHS & IHS \\
        K2\hyp{}265 b & $0.24\pm0.24$ & $0.33\pm0.07$ & 94  & IHS & IHS \\
        HD 213885 b & $0.42\pm0.09$ & $0.31\pm0.07$ & 66  & IHS & IHS  \\  
        WASP\hyp{}47 e & $0.155^{+0.14}_{-0.15}$ & $0.26\pm 0.07$  & 80  & IHS & IHS \\
        Kepler\hyp{}20 b & $0.26^{+0.14}_{-0.16}$ & $0.30\pm0.10$ & 98 & IHS & IHS \\
        55 Cnc e & $0.004^{+0.10}_{<0}$ & $0.31\pm0.10$ & 9  & LDSP & IHS  \\

    \hline
    \end{tabular}
    \caption{Inferred properties of the selected small, well\hyp{}characterized exoplanets in order of increasing radius. IHS: Indistinguishable from Host Star, SM: super\hyp{}Mercury, or LDSP: Low\hyp{}Density Small Planet. While we classify planets at the $1\sigma$ level, we only consider planets that deviate from their host stars at the $2 \sigma$ level to be conclusively \textit{inconsistent} with the null hypothesis. The CMF$_\rho$ and $P(\mathcal{H}^0)$ values from earlier published M and R are in Table \ref{tab:prev_studies}. Planets are listed in order of increasing radius as in Table \ref{table:SAMPLE}.}
    \label{table:RESULTS}

\end{table}

\section{Cases for a Rejected Null Hypothesis}

\vspace{0.25 cm}
\noindent For planets that deviate from $\mathcal{H}^0$ in Table \ref{table:RESULTS}, there are two cases, (1) in which CMF$_{\mathrm{star}}$ $<$ CMF$_\rho$, suggesting a planet with a larger core than expected, a super\hyp{}Mercury, and (2) in which CMF$_{\mathrm{star}}$ $>$ CMF$_\rho$, a low\hyp{}density small planet, which then suggests a region of non\hyp{}unique solutions. We define and use the term Low\hyp{}Density Small Planet (LDSP) for planets with an apparent density deficit, but distinguished from super\hyp{}Puffs, a class of super\hyp{}Earth mass planets with gas\hyp{}giant transit radii leading to bulk densities of $\lesssim 0.1$ g/cc \citep[e.g.,][and references therein]{superPuff}. In contrast, LDSPs have bulk densities that can only be explained with a rock\hyp{}dominated composition but sufficiently low that they are inconsistent with $\mathcal{H}^0$. For example, the candidate LDSP 55 Cnc e has a mean bulk density of $\sim 6.25$ g/cc which, for its mass, is consistent with a rocky (pure MgSiO$_3$) composition, but is $\sim 1.76$ g/cc lower than its expected bulk density per $\mathcal{H}^0$.

\vspace{0.25 cm}
\noindent
Kepler\hyp{}107 c and 55 Cnc e, the two planets inconsistent with $\mathcal{H}^0$ at $>$ 1$\sigma$ level, differ in that Kepler 107c has a mass excess relative to what would be predicted by its host star, while 55 Cnc e has a relative mass deficit. This demands multiple explanations for these planets and suggests a diversity of planetary outcomes for planets that orbit close to their stars.

\subsection{Super\hyp{}Mercuries}
\label{SMs}

\vspace{0.25 cm}
\noindent
Kepler-107c has a greater core mass fraction than predicted by its host star's Fe/Mg abundances. There are few explanations for increasing the density of a planet beyond excess iron relative to MgO and SiO$_2$. The simplifying assumption of a solid, pure iron core means that $P(\mathcal{H}^0$) is an upper bound. Therefore, this planet is likely a super\hyp{}Mercury. 

\vspace{0.25 cm}
\noindent
Considering the actual core properties of the rocky solar system planets, it is likely that the cores of exoplanets contain some amount of alloying light elements and are at least partially liquid. The cores of the terrestrial solar system planets contain $\sim$ 10\% light elements and are volumetrically dominated by liquid \citep[e.g.,][]{Mars_core_and_LEs, Merc_evidence_for_liq_outer_core, Venus_int_comp, Birch_LE_enrich, Lehmann_Earths_Core}. Including these factors only strengthens the case for a large core for Kepler-107c. 

\vspace{0.25 cm}
\noindent
Both light element incorporation and liquid iron reduce the density of the core. For models that conserve both mass and radius, our simplifying assumption of overestimating the core density underestimates core volume due to systematically bounding the density as a likely maximum. CMF$_\rho$ values assuming a pure and solid Fe core are $\sim$ 0.02\hyp{}0.04 lower than the liquid and light-element-enriched core cases for a given planetary mass and radius. As an example, assuming a liquid core for K2\hyp{}229 b, the second most probable super\hyp{}Mercury, increases CMF$_\rho$ by $\sim$0.02 relative to a solid core, corresponding to a 5\% decrease in $P(\mathcal{H}^0)$.

\vspace{0.25 cm}
\noindent
An additional assumption is that all iron is in the core and none in the mantle. In the calculation of CMF$_{\mathrm{star}}$, Fe/Mg is constant, but the oxidation of iron removes Fe from the core as well as adds oxygen to the planet, decreasing the effective CMF$_{\mathrm{star}}$ by 0.02\hyp{}0.03 compared to the FeO\hyp{}free case. The comparable CMF$_\rho$ calculation, however, oxidizes Fe from the core incorporating it into the mantle as FeO. This increases the average density of the mantle while decreasing the density of the planet through added oxygen, decreased compressibility of the oxide compared to the metal, and decreased core mass. For example, an Earth-like, whole-rock assemblage has 4 mol \% FeO, accounting for oxidation of 13 mol\% of Earth's iron, has a bulk modulus of ~250 GPa \citep{KKMLee04}, while the bulk modulus of solid, hcp iron is 178 GPa \citep{Smith18_Fe_EoS}. As a result, the simplifying assumption that all iron is in the core, a resulting model for CMF$_\rho$ at a given mass and radius is an underestimate for planet iron fraction by $\sim$ 0.01\hyp{}0.04. For example, the assumption of a Fe\hyp{}free mantle in K2\hyp{}229 b predicts CMF$_\rho=0.565$. 10 mol\% oxidation of the available iron, removing it from the core and incorporating it into the mantle as an iron oxide, requires a 0.032 increase in planet Fe/Mg for K2\hyp{}229 b at its given mass to reproduce this planet's radius. Given constant stellar Fe/Mg, this increase in iron corresponds to an 8\% decrease in $P(\mathcal{H}^0)$ relative to the Fe\hyp{}free mantle case.

\vspace{0.25 cm}
\noindent For both assumptions of a solid, pure iron core and iron\hyp{}free mantle, the probability assigned to $\mathcal{H}^0$ is an upper bound in the case CMF$_\rho$ $>$ CMF$_{\mathrm{star}}$, and, therefore, the 5\% probability criterion for super\hyp{}Mercuries in this situation is a conservative measure.

\vspace{0.25 cm}
\subsection{Low\hyp{}Density Small Planets}

\vspace{0.25 cm}
\noindent
Four compositional variations can potentially explain LDSPs, (1) significant oxidization of iron such that it is removed from the core and incorporated into the mantle (e.g. \citet{Rogers10_FeSiO3}), (2) a calcium\hyp{}aluminum oxide dominated planet that formed from only these highest\hyp{}temperature refractory materials with significantly depleted Fe (e.g. \citet{Dorn_CAI_Planets}), (3) volatile outer layers (e.g. \citet{ehrenreich_volatile, Crida18_55Cnce, tsiaras16_55Cnce, angelo17_55Cnce, dorn17_bayes}), or (4) or significant melt fraction \citep{Bower19MagmaOceanPlanets}. 

\begin{figure}[ht]
    \centering
    \includegraphics[scale = 0.5]{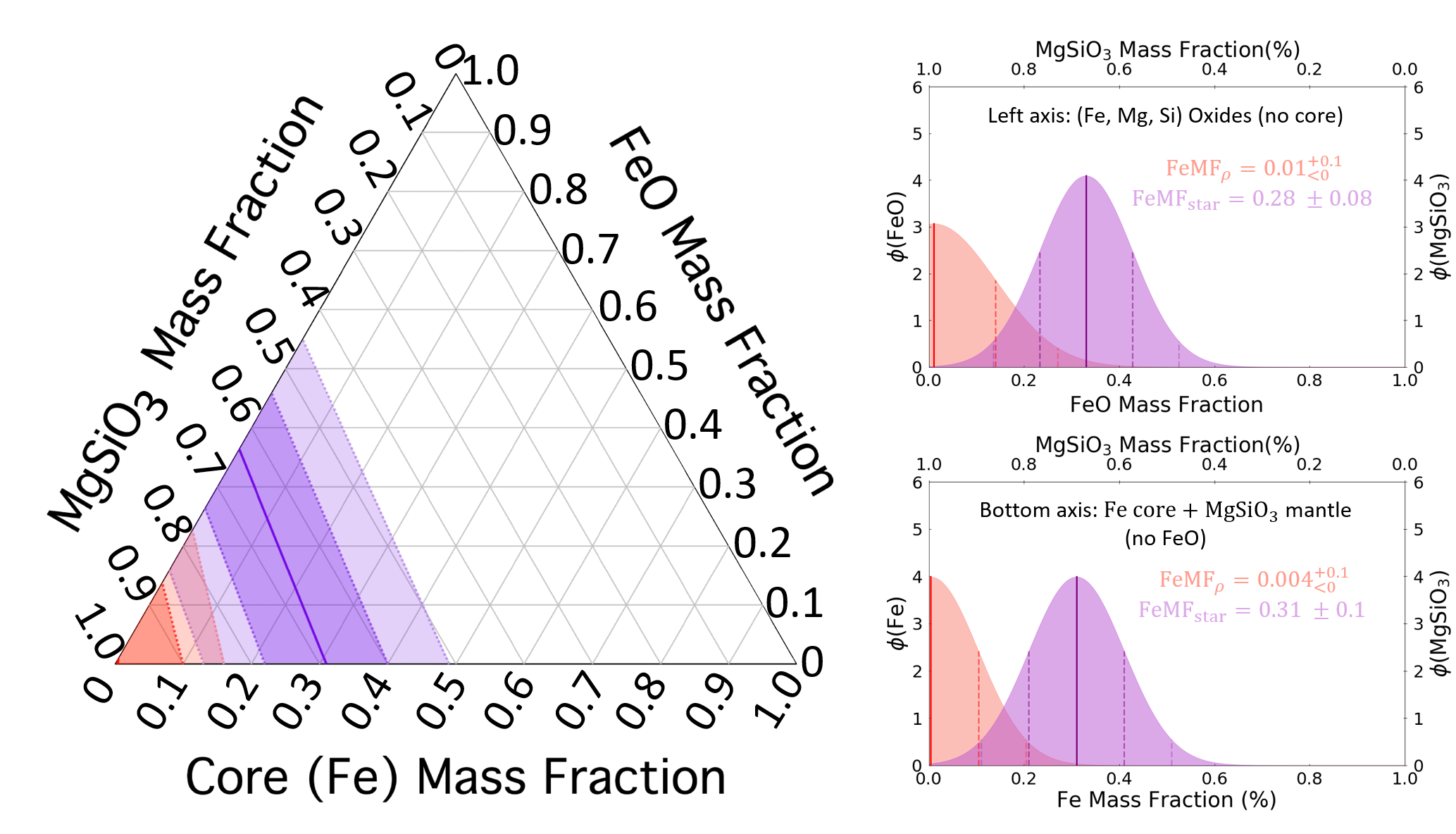}
    \caption{Ternary diagram of the CMF\hyp{}MMF\hyp{}FeO solution space for 55 Cnc e. All possible CMF$_{\mathrm{star}}$ solutions  (purple, dark 1$\sigma$, light 2$\sigma$) and CMF$_\rho$ solutions (red, dark 1$\sigma$, light 2$\sigma$) are plotted with the constraint that CMF+MgSiO$_3$+FeO $= 100\%$. Our model assumes that the core is pure Fe, and all (Fe, Mg, Si) oxides reside in the mantle. FeMF$_\rho$ and FeMF$\mathrm{_{star}}$ refer to the bulk Fe mass fraction inferred from mass\hyp{}radius measurements and host\hyp{}star abundances, respectively. In the case of no oxidation, FeMF$_\rho$ = CMF$_\rho$ and FeMF$\mathrm{_{star}}$ = CMF$_{\mathrm{star}}$.}
    \label{fig:55Cnce_tern}
\end{figure}

\vspace{0.25 cm}
\noindent
As with the impact of oxidation on K2\hyp{}229 b, we explore the degree to which core oxidation affects our analysis of the null hypothesis in the case of LDSPs. In the case that 55 Cnc e is core\hyp{}free due to oxidation of all iron, $P(\mathcal{H}^0)$ increases from 9\% to 10\% (Fig. \ref{fig:55Cnce_tern}). As in \ref{SMs}, this is a consequence of the iron being incorporated in a lower density, lower compressibility oxide as compared to the metal. At the same time, this results in an associated decrease in CMF$_{\mathrm{star}}$ due to an added oxygen atom per iron atom. 

\vspace{0.25 cm}
\noindent
A major objection to the possibility of a planet made of ultra\hyp{}high temperature condensates as proposed by \citet{Dorn_CAI_Planets} is that there is an insufficient mass of Al and Ca present within the protoplanetary disk available to produce the observed masses of these planets. For instance, reproducing the mass of 55 Cnc e assuming it formed from a minimum mass solar nebula \cite{Kuch04}, which likely overestimates the disk mass available to planets forming around its K\hyp{}dwarf host, requires a factor of $\sim$3.5 (0.55 dex) and $\sim$2 (0.3 dex) increase in the already super\hyp{}solar Ca and Al abundances of 55 Cnc, respectively. That being said, the mean density of 55 Cnc e is consistent with a virtually iron\hyp{}free planet of MgSiO$_3$, similar to the compositional prediction in this hypothesis.

\vspace{0.25 cm}
\noindent
\noindent Where the lower\hyp{}than\hyp{}expected density cannot be explained by oxidation or iron deficit, the most likely explanation is a combination of H/He or higher\hyp{}mass atmospheric compositions including H$_2$O and CO$_2$ (e.g. \citet{ehrenreich_volatile, Crida18_55Cnce, tsiaras16_55Cnce, angelo17_55Cnce, dorn17_bayes}). For 55 Cnc e, the radius deficit between what is observed and what is expected per $\mathcal{H}^0$ can be explained by a $\sim1000$ km thick atmosphere. Given the proximity of 55 Cnc e to its host, an H/He or H$_2$O-dominated atmosphere would result in escaping hydrogen which has not been observed \citep[e.g.,][and references therein]{Bourrier18_55Cnce}. A water-rich atmosphere has recently been ruled out at the 3$\sigma$ level \citep{Jindal2020}: If 55 Cnc e does have an atmosphere, it must be dominated by heavier species including CO, CO$_2$, or N$_2$.

\vspace{0.25 cm}
\noindent We note that earlier, larger, measurements of the radius of 55 Cnc e and WASP\hyp{}47e place these planets as inconsistent with $\mathcal{H}^0$ at greater than 2$\sigma$ and 1$\sigma$ level, respectively (Table \ref{tab:prev_studies}), which would predict both planets as LDSPs. Both planet radii have been revised downward using updated stellar parameters from {\it Gaia} parallaxes (\citet{Dai_Homogenous_MRs_HotEarths}).

\subsection{Necessary Observational Improvements to Reject the Null Hypothesis}

\vspace{0.25 cm}
\noindent
Current efforts are underway to improve mass and radius measurements for rocky exoplanets. We investigate here the needed improvements in the observational uncertainties of mass, radius, and host\hyp{}star abundances to help quantify the range of rocky planet compositions. We find that increasing precision in planetary radius measurements is the most critical.

\begin{figure}[ht]
    \centering
    \includegraphics[scale = 0.4]{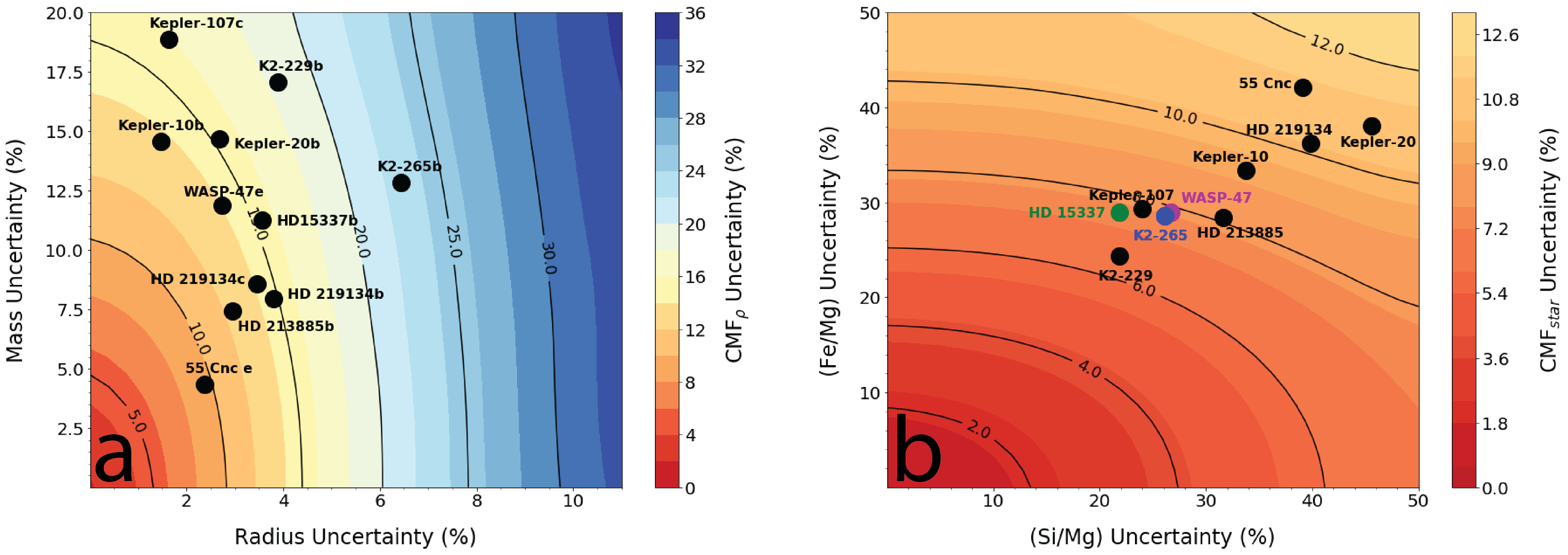}
    \caption{(a) Uncertainty in CMF$_\rho$ as a function of planet mass and radius uncertainties. Uncertainties here assume $M_p = 5.0 M_\oplus$ and $R_p = 1.545 R_\oplus$ for a central value for CMF$_\rho$=0.35. Values in Table \ref{table:RESULTS} are exact. (b) Uncertainty in CMF$_{\mathrm{star}}$ as a function of uncertainties in the stellar, molar Fe/Mg and Si/Mg. We plot uncertainties for central values of Fe/Mg = 0.98 and Si/Mg = 1.0 which result in a central value of 0.35 for  CMF$_{\mathrm{star}}$. HD 15337, K2\hyp{}265, and WASP\hyp{}47 are colored for clarity.}
    \label{fig:heatmaps}
\end{figure}

\vspace{0.25 cm}
\noindent
Across the radius and mass range of the analysis (2.5\hyp{}9.7 $M_\oplus$ and 1.1\hyp{}1.9 $R_\oplus$), we calculate the relative effects of uncertainties in measured mass, radius, and stellar abundances in CMF$_\rho$ and CMF$_{\mathrm{star}}$ (Figure \ref{fig:heatmaps}). Uncertainties arising in mass and/or radius resulting from the uncertainties in the underlying equation of states for each layer are minimal \citep{unterborn_panero19}. The uncertainty in CMF$_\rho$ is weakly dependent upon the planetary mean density in the considered range (Table \ref{dcmfrhotab}).

\vspace{0.25 cm}
\noindent
For a typical planet with CMF$_\rho$ = 0.35, we determine that a 20\% uncertainty in mass for a planet leads to a CMF$_\rho$ uncertainty of 0.15 (Figure \ref{fig:heatmaps}a, Table \ref{dcmfrhotab}). The observational uncertainties in planet radius have the greatest impact on inferred CMF$_\rho$. A 10\% radius uncertainty, again for a typical CMF$_\rho$ of 0.35, leads to an uncertainty that is as large as the central value, i.e. CMF = 0.35 $\pm$ 0.35. 

\vspace{0.25 cm}
\noindent
The observational uncertainty in host\hyp{}star Fe, Mg, and Si abundances have a proportionally smaller effect on CMF$_{\mathrm{star}}$ (Table \ref{dcmfstar}). A 40\% uncertainty in molar Fe/Mg leads to a CMF$_{\mathrm{star}}$ uncertainty $<$ 0.10 (Figure \ref{fig:heatmaps}b). We find that the uncertainty in molar Si/Mg has minimal impact. 

\vspace{0.25 cm}
\noindent
Improving mass and radius uncertainties for small planets is a resource\hyp{}intensive process, which will benefit from increased signal\hyp{}to\hyp{}noise ratios of {\it TESS} \citep{tess} and {\it CHEOPS} \citep{CHEOPS18} for bright hosts, along with parallax measurements from {\it Gaia} \citep{gaia}, which may improve mean planet density observational uncertainty to as little as 4\% \citep{Stevens18_MR_precision}. From the stellar perspective, direct analysis for [Fe/Mg] and [Si/Mg], circumvents compounding effects of covariances and permits for a reduction in uncertainties relative to calculating abundances from [Fe/H], [Mg/H], and [Si/H] \citep{Epstein10}.

\vspace{0.25 cm}
\noindent

\begin{figure}[ht]
    \centering
    \includegraphics[scale = 0.75]{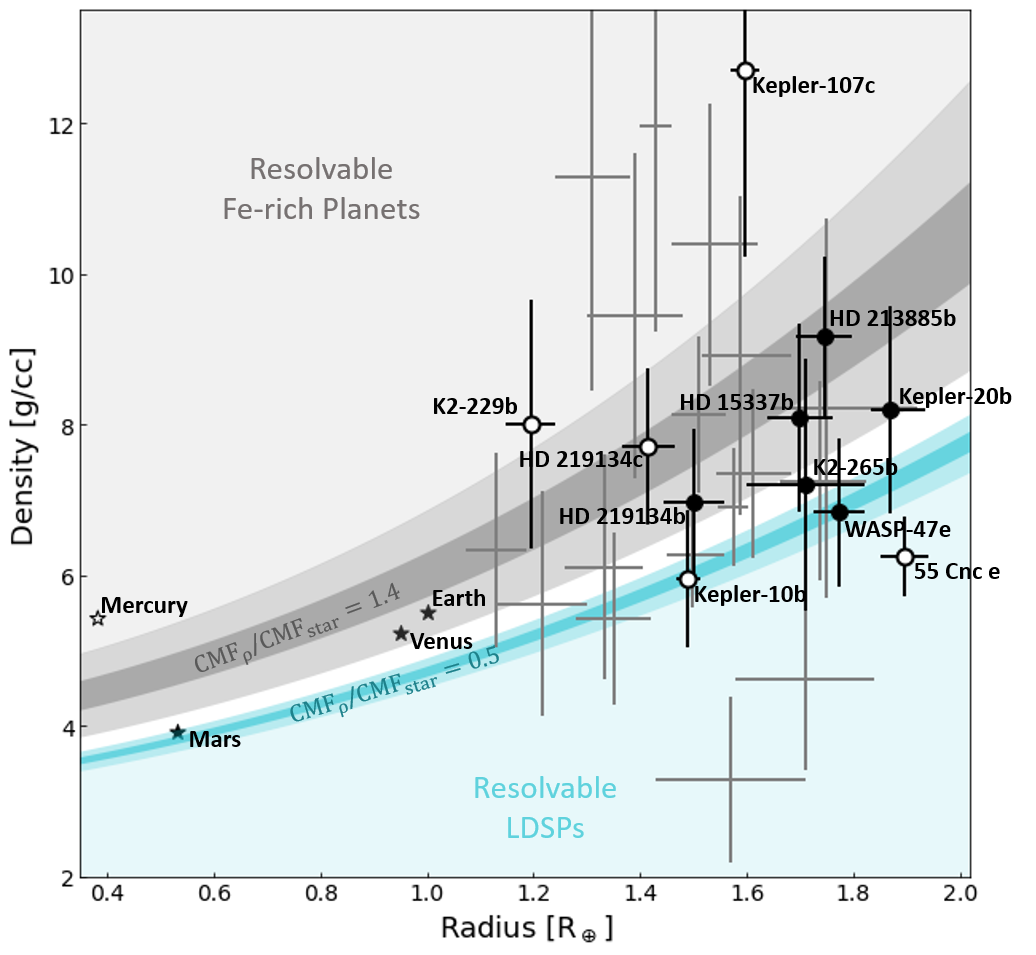}
    
    \caption{Planet density as a function of radius for the 11 planets in our sample (circles) and the solar system terrestrial planets (stars). The observational radius and density uncertainties are plotted for each planet. Open circles indicate planets that may be distinguishable from $\mathcal{H}^0$ at the 2$\sigma$ significance level should $\sigma_{M_p} = 4\%$, $\sigma_{R_p} = 1\%$, and $\sigma_{\mathrm{Fe/Mg}} =  8\%$ precisions be attained. The dark\hyp{}gray shaded region covers the density range of our sample for which $\mathrm{CMF_{\rho}}$/$\mathrm{CMF_{star}}$ = 1.4, while the light gray shading reflects the middle 95\% range of $\mathrm{CMF_{star}}$ from the stellar abundances in the Hypatia catalog. Similarly, the cyan shading reflects the range of this sample and of the equivalent Hypatia range where $\mathrm{CMF_{\rho}}$/$\mathrm{CMF_{star}}$ = 0.5. Gray crosses are the 17 planets that meet our mass and radius criteria but lack Fe, Mg, and Si abundances measurements of their host stars. These planets are listed in Table \ref{table:AllpSAMPLE}.}
    \label{fig:Des_v_Rp_summary_plot}
\end{figure}

\vspace{0.25 cm}
\noindent
Assuming a near best\hyp{}case precision of 4\% in mass, 1\% in radius (5\% $\rho$ uncertainty), and 8\% in Fe/Mg and Si/Mg stellar abundances (corresponding to $\sim$ solar Fe/Mg and Si/Mg uncertainties), for planets in the range of 2.5\hyp{}9.7 $M_\oplus$ and 1.1\hyp{}1.9 $R_\oplus$, the null hypothesis cannot be refuted when 0.5 $<$ CMF$_\rho$/CMF$_{\mathrm{star}}$ $<$ 1.4. For the planets in our sample set, assuming accurate central values for each measurement, Kepler\hyp{}10b and 55 Cnc e may be conclusively determined to have lower\hyp{}than\hyp{}expected densities, while HD 219134c and K2\hyp{}229 b may be sufficiently resolved as planets with greater\hyp{}than\hyp{}expected densities, or super\hyp{}Mercuries (Fig. \ref{fig:Des_v_Rp_summary_plot}, Table \ref{tab:min_unc_table}). The remaining planets, HD 219134 b, K2\hyp{}265b, WASP\hyp{}47e, HD 15337 b, HD 213885b, and Kepler 20b cannot be distinguished from the null hypothesis at the 2$\sigma$ level using mass, radius, and stellar abundances.

\vspace{0.25 cm}
\noindent
Next, we consider Mercury, Earth, Mars. While these planets are smaller than those in our sample, their geophysical constraints on CMF provide test cases for the validity of our methods and assumptions. MESSENGER constrained Mercury's CMF to $\sim0.69 \hyp{} 0.77$, corresponding to CMF$_\rho$/CMF$_{\mathrm{star}} = 2.17 - 2.43$ \citep[][]{nittler}. Were Mercury to be viewed as an exoplanet with near-best case observation precisions, it should, therefore, be resolvable as denser than expected. Applying the methods and assumptions outlined in this work, we indeed find Mercury would be resolvable as denser than expected with CMF$_\rho$/CMF$_{\mathrm{star}} = 1.92$ and $P(\mathcal{H}^0)$ indistinguishable from zero. Earth has a CMF$_\rho$/CMF$_{\mathrm{star}}$ of 1.02 \citep[][and references therein]{McDonough2017}. Thus, even with best-case observational precisions, the Earth as an exoplanet would be indistinguishable from the Sun. Using our approach on Earth, we find CMF$_\rho$/CMF$_{\mathrm{star}}$ = 1.03 and $P(\mathcal{H}^0)$ = 99\%. Last, while Mars has a molar Fe/Mg ratio to within $\sim10\hyp{}15\%$ of the Sun's, much of its Fe is oxidized, leading to a smaller CMF than the Earth. Geophysical constraints give CMF$_\rho$/CMF$_{\mathrm{star}} \sim 0.57-0.67$ for Mars. Thus, even with significant oxidation of iron, Mars as an exoplanet should still be indistinguishable from the Sun at the 2$\sigma$  level. Our approach suggest values on the lower end without iron oxidation,  CMF$_\rho$/CMF$_{\mathrm{star}}$ = 0.57 and $P(\mathcal{H}^0)$ = 11\% \citep{Wanke94_Chem_and_Acc_Hist_Mars, Bertka_Fei_1998, Zharkov_Gudkova_2005, Yoshizaki_McDonough20}.

\section{Discussion and Conclusions}

\vspace{0.25 cm}
\noindent We assess the statistical consistency of a planet's composition and structure as inferred from its mass and radius with what is expected from its host's major rock\hyp{}building elemental abundance ratios. We test the hypothesis directly and demonstrate that for just one planet, Kepler 107c, the mass and radius cannot be described as a terrestrial planet with the same relative major rock\hyp{}building element abundances as its host, demanding a two\hyp{}fold excess of iron. This approach is complementary to the Bayesian approach in \citet{Dorn15_A&A_bayesian_rocky_interiors} and \citet{Otegi20_preprint}, which leverage stellar abundances to reduce planetary structure degeneracy, in which the null hypothesis tested here is assumed a priori. Once the conditions under which a terrestrial planet is describable by its host's abundances are understood, a more complete Bayesian analysis of planetary composition and structure will be warranted.

\vspace{0.25 cm}
\noindent 
We demonstrate that super\hyp{}Mercuries with an iron overabundance that is at least 40\% greater than CMF$_{\mathrm{star}}$ may be resolvable as having different from host\hyp{}star major rock\hyp{}building element compositions. For example, we show that HD 219134c will be a resolvable super\hyp{}Mercury while HD 213885b will remain unresolvable despite these planets having the same mean $\mathrm{CMF}_\rho$. The difference arises from HD 213885 being more Fe\hyp{}rich relative to silicates than HD 219134 leading to 135\% and 150\% iron overabundances for HD 213885b and HD 219134b, respectively. 

\vspace{0.25 cm}
\noindent
Several planetary formation and evolution mechanisms may explain super\hyp{}Mercury planets. Each model identifies mechanisms by which planets are enriched in iron relative to their host star. These models include giant impacts \citep{Leinhardt_2011, Marcus10_MinRad_SEs}, a series of smaller impacts \citep{Chau_2018, Swain19_NewClass_Terrestrial_Planets}, mantle evaporation of hot planets \citep[e.g.,][and references therein]{Santerne18_K2229b}, iron\hyp{}enrichment in the inner regions of planet\hyp{}forming disks due to iron condensing at a higher temperature than silicate material \citep{LEWIS1972286} or via photophoresis \citep[e.g.,][and references therein]{ebel2019}, and mantle stripping via planet\hyp{}star tidal interactions \citep{Shi16_mass_transfer} or planet\hyp{}planet tidal interactions \citep{Deng20_Planet-Planet_tidal_stripping}. Each model to explain the formation of Mercury or super\hyp{}Mercuries involves compositional sculpting of such planets relative to their host star. The frequency and orbital properties of exo\hyp{}super\hyp{}Mercuries are a crucial test of these theories.

\vspace{0.25 cm}
\noindent
HD 219134c is potentially resolvable as a super\hyp{}Mercury only when analyzed relative to its host's abundances. HD219134 is a star with proportionally little Fe with Fe/Mg = 0.69 $\pm$ 0.25, yet slightly iron\hyp{}enriched relative to Solar, with [Fe/H] = 0.09 \citep{Hinkel14_Hypatia}. Had the planetary structure analysis focused solely on the CMF$_\rho$ ($0.42^{+0.13}_{-0.14}$), this planet may have been missed as a planet that underwent significant chemical sculpting, as it is less than 1$\sigma$ greater than an Earth\hyp{}like CMF of 0.32. Notably, this planet orbits outside that of HD 219134 b, which is indistinguishable from its host star. Similarly, K2\hyp{}229 b may have been prematurely misidentified as a super\hyp{}Mercury, even with the updated mass and radius values used in this study, had its host star's abundances not been considered. K2\hyp{}229 b and HD 219134 c show that stellar major rock\hyp{}building element composition \text{must} be carefully considered when investigating the outcome diversity of rocky planet formation to avoid failing to or misidentifying a planet as a super\hyp{}Mercury or LDSP.

\vspace{0.25 cm}
\noindent
We also show that we can identify low\hyp{}density, small planets whose CMF$_\rho$ is 50\% than predicted by CMF$_\mathrm{star}$. The source of the low density is degenerate, but the 1$\sigma$ mass deficit for 55 Cnc e cannot be explained through oxidation of all iron. The relative influence of atmospheric layers, a dramatic iron deficit, or global magma oceans remain degenerate. Such planets offer important clues as to planetary system evolution, demonstrating that small planets below the radius gap cannot be exclusively rocky planets.

\vspace{0.25 cm}
\noindent
The approach of comparing expected density based on host\hyp{}star composition to that of the planet only addresses the most extreme cases of compositional sculpting. The remaining planets in the sample set, more than half, are not distinguishable from the null hypothesis, nor will they be distinguishable from their host star with respect to composition based on mass and radius measurements alone. For potential LDSPs, further constraint on the differences between host\hyp{}star composition and rocky planet composition must come from alternate methods. First, the measurement of day-to-night side temperature contrast could reveal a thick atmosphere and thus a LDSP~\citep{Koll19}. Second, the measurement of planet atmospheric composition and relative abundance ratio~\citep{Morley_JWST_atmochar_17} can be compared to that of the host star.

\vspace{0.25 cm}
\noindent
\citet{Plotnykov20} and \citet{Scora20} have recently presented complementary approaches to detecting rocky planets with non-stellar compositions. \citet{Plotnykov20} compare the mean planetary CMF distribution ($\langle\mathrm{CMF}_\rho\rangle$) for likely rocky planets with $\mathrm{\sigma_{M_p}}$ and $\mathrm{\sigma_{R_p}}$ $<$ 25\% to the mean stellar CMF distribution ($\langle\mathrm{CMF_{star}}\rangle$) of the stars in the Hypatia Catalog \citep{Hinkel14_Hypatia}. The authors find that the $\mathrm{CMF}_\rho$ distribution peaks at a lower value and is more broad than the $\mathrm{CMF_{star}}$ distribution, requiring an explanation in planet formation theories. \citet{Scora20} investigate whether the diversity of $\langle\mathrm{CMF}_\rho\rangle$ relative to what is expected from $\langle\mathrm{CMF_{star}}\rangle$ can be explained through the cumulative effects of collisions during formation, finding that collisions alone cannot explain the diversity of rocky planet compositions.

\vspace{0.25 cm}
\noindent
While comparing statistical average values for CMF$_\rho$  and $\mathrm{CMF_{star}}$ is useful for assessing planets in which no stellar abundance measurements are made, it fails to consider the composition deviation in Fe/Mg from star to planet on an individual basis. Assuming a best case CMF$_\rho$ uncertainty of 0.03 (Kepler-107c Table \ref{tab:min_unc_table}) and $\langle\mathrm{CMF_{star}}\rangle \; = 0.32^{+0.14}_{-0.12}$ \citep{Plotnykov20}, only planets with CMF$_\rho <0.02$ or $>0.67$ (ratios of $<0.06$ and $>2.1$, respectively) may be resolvable as statistically inconsistent with $\langle\mathrm{CMF_{star}}\rangle$.  As observational precisions increase and Fe/Mg and Si/Mg values become more widely reported, our approach will be able to identify these same compositional extremes and more  modest  cases  of  compositional sculpting.

\vspace{0.25 cm}
\noindent To assess  the  formation  processes  that  result  in  measurable compositional deviations for planets  smaller than the radius gap will require a larger, more precise sample with both extreme and more moderate cases of compositional sculpting. In our sample selection process, we identified 17 small planets with high\hyp{}precision mass and radius measurements, but without host\hyp{}star abundance measurements beyond [Fe/H] (Table \ref{table:AllpSAMPLE}). The sample\hyp{}set, therefore, may be doubled rapidly with targeted, high\hyp{}precision stellar abundance analysis. With a larger data set, deviations from the expected can be addressed statistically much like the radius populations are able to do now. It is \textit{imperative} that host\hyp{}star Fe, Mg, and Si abundance measurements are included along with mass and radius measurements of likely rocky planets as part of the discussion and inference on their structure and composition. Without mass\hyp{}radius \textit{and} host\hyp{}star abundance measurements, it is impossible to determine if a planet's composition reflects that of its host and, in turn, determine its most likely formation pathways.

\acknowledgments
\noindent 
CTU acknowledges the support of Arizona State University through the SESE Exploration fellowship. The results reported herein benefited from collaborations and/or information exchange within NASA's Nexus for Exoplanet System Science (NExSS) research coordination network sponsored by NASA's Science Mission Directorate. JGS acknowledges the support of The Ohio State School of Earth Sciences through the Friends of Orton Hall research grant. We acknowledge partial support to WRP from the National Science Foundation under Grant No. EAR\hyp{}1724693. This research has made use of the NASA Exoplanet Archive, which is operated by the California Institute of Technology, under contract with the National Aeronautics and Space Administration under the Exoplanet Exploration Program.

\bibliography{biblio}{}

\begin{thebibliography}{}
\expandafter\ifx\csname natexlab\endcsname\relax\def\natexlab#1{#1}\fi
\providecommand{\url}[1]{\href{#1}{#1}}
\providecommand{\dodoi}[1]{doi:~\href{http://doi.org/#1}{\nolinkurl{#1}}}
\providecommand{\doeprint}[1]{\href{http://ascl.net/#1}{\nolinkurl{http://ascl.net/#1}}}
\providecommand{\doarXiv}[1]{\href{https://arxiv.org/abs/#1}{\nolinkurl{https://arxiv.org/abs/#1}}}

\bibitem[{{Aitta}(2012)}]{Venus_int_comp}
{Aitta}, A. 2012, \icarus, 218, 967, \dodoi{10.1016/j.icarus.2012.01.007}

\bibitem[{{Almenara} {et~al.}(2016){Almenara}, {D{\'\i}az}, {Bonfils}, \&
  {Udry}}]{Almenara_WASP47e}
{Almenara}, J.~M., {D{\'\i}az}, R.~F., {Bonfils}, X., \& {Udry}, S. 2016, \aap,
  595, L5, \dodoi{10.1051/0004-6361/201629770}

\bibitem[{{Angelo} \& {Hu}(2017{\natexlab{a}})}]{AngeloHu_55Cnce}
{Angelo}, I., \& {Hu}, R. 2017{\natexlab{a}}, \aj, 154, 232,
  \dodoi{10.3847/1538-3881/aa9278}

\bibitem[{{Angelo} \& {Hu}(2017{\natexlab{b}})}]{angelo17_55Cnce}
---. 2017{\natexlab{b}}, \aj, 154, 232, \dodoi{10.3847/1538-3881/aa9278}

\bibitem[{{Astudillo-Defru} {et~al.}(2020){Astudillo-Defru}, {Cloutier},
  {Wang}, {Teske}, {Brahm}, {Hellier}, {Ricker}, {Vand erspek}, {Latham},
  {Seager}, {Winn}, {Jenkins}, {Collins}, {Stassun}, {Ziegler}, {Almenara},
  {Anderson}, {Artigau}, {Bonfils}, {Bouchy}, {Brice{\~n}o}, {Butler},
  {Charbonneau}, {Conti}, {Crane}, {Crossfield}, {Davies}, {Delfosse},
  {D{\'\i}az}, {Doyon}, {Dragomir}, {Eastman}, {Espinoza}, {Essack}, {Feng},
  {Figueira}, {Forveille}, {Gan}, {Glidden}, {Guerrero}, {Hart}, {Henning},
  {Horch}, {Isopi}, {Jenkins}, {Jord{\'a}n}, {Kielkopf}, {Law}, {Lovis},
  {Mallia}, {Mann}, {de Medeiros}, {Melo}, {Mennickent}, {Mignon}, {Murgas},
  {Nusdeo}, {Pepe}, {Relles}, {Rose}, {Santos}, {S{\'e}gransan}, {Shectman},
  {Shporer}, {Smith}, {Torres}, {Udry}, {Villasenor}, {Winters}, \&
  {Zhou}}]{AD20_L1689b_MR}
{Astudillo-Defru}, N., {Cloutier}, R., {Wang}, S.~X., {et~al.} 2020, \aap, 636,
  A58, \dodoi{10.1051/0004-6361/201937179}

\bibitem[{{Batalha} {et~al.}(2011){Batalha}, {Borucki}, {Bryson}, {Buchhave},
  {Caldwell}, {Christensen-Dalsgaard}, {Ciardi}, {Dunham}, {Fressin},
  {Gautier}, {Gilliland}, {Haas}, {Howell}, {Jenkins}, {Kjeldsen}, {Koch},
  {Latham}, {Lissauer}, {Marcy}, {Rowe}, {Sasselov}, {Seager}, {Steffen},
  {Torres}, {Basri}, {Brown}, {Charbonneau}, {Christiansen}, {Clarke},
  {Cochran}, {Dupree}, {Fabrycky}, {Fischer}, {Ford}, {Fortney}, {Girouard},
  {Holman}, {Johnson}, {Isaacson}, {Klaus}, {Machalek}, {Moorehead},
  {Morehead}, {Ragozzine}, {Tenenbaum}, {Twicken}, {Quinn}, {VanCleve},
  {Walkowicz}, {Welsh}, {Devore}, \& {Gould}}]{Batalha_kep10b}
{Batalha}, N.~M., {Borucki}, W.~J., {Bryson}, S.~T., {et~al.} 2011, \apj, 729,
  27, \dodoi{10.1088/0004-637X/729/1/27}

\bibitem[{{Benz} {et~al.}(2018){Benz}, {Ehrenreich}, \& {Isaak}}]{CHEOPS18}
{Benz}, W., {Ehrenreich}, D., \& {Isaak}, K. 2018, {CHEOPS: CHaracterizing
  ExOPlanets Satellite}, 84, \dodoi{10.1007/978-3-319-55333-7_84}

\bibitem[{{Bertka} \& {Fei}(1998)}]{Bertka_Fei_1998}
{Bertka}, C.~M., \& {Fei}, Y. 1998, Earth and Planetary Science Letters, 157,
  79, \dodoi{10.1016/S0012-821X(98)00030-2}

\bibitem[{{Birch}(1952)}]{Birch_LE_enrich}
{Birch}, F. 1952, \jgr, 57, 227, \dodoi{10.1029/JZ057i002p00227}

\bibitem[{{Bonfils} {et~al.}(2018){Bonfils}, {Almenara}, {Cloutier},
  {W{\"u}nsche}, {Astudillo-Defru}, {Berta-Thompson}, {Bouchy}, {Charbonneau},
  {Delfosse}, {D{\'\i}az}, {Dittmann}, {Doyon}, {Forveille}, {Irwin}, {Lovis},
  {Mayor}, {Menou}, {Murgas}, {Newton}, {Pepe}, {Santos}, \&
  {Udry}}]{Bonfils18_GJ1132MR}
{Bonfils}, X., {Almenara}, J.~M., {Cloutier}, R., {et~al.} 2018, \aap, 618,
  A142, \dodoi{10.1051/0004-6361/201731884}

\bibitem[{{Bonomo} {et~al.}(2019){Bonomo}, {Zeng}, {Damasso}, {Leinhardt},
  {Justesen}, {Lopez}, {Lund}, {Malavolta}, {Silva Aguirre}, {Buchhave},
  {Corsaro}, {Denman}, {Lopez-Morales}, {Mills}, {Mortier}, {Rice}, {Sozzetti},
  {Vanderburg}, {Affer}, {Arentoft}, {Benbakoura}, {Bouchy},
  {Christensen-Dalsgaard}, {Collier Cameron}, {Cosentino}, {Dressing},
  {Dumusque}, {Figueira}, {Fiorenzano}, {Garc{\'\i}a}, {Hand berg},
  {Harutyunyan}, {Johnson}, {Kjeldsen}, {Latham}, {Lovis}, {Lundkvist},
  {Mathur}, {Mayor}, {Micela}, {Molinari}, {Motalebi}, {Nascimbeni}, {Nava},
  {Pepe}, {Phillips}, {Piotto}, {Poretti}, {Sasselov}, {S{\'e}gransan}, {Udry},
  \& {Watson}}]{Bonomo19_Kep107_Nature}
{Bonomo}, A.~S., {Zeng}, L., {Damasso}, M., {et~al.} 2019, Nature Astronomy, 3,
  416, \dodoi{10.1038/s41550-018-0684-9}

\bibitem[{{Bourrier} {et~al.}(2018){Bourrier}, {Dumusque}, {Dorn}, {Henry},
  {Astudillo-Defru}, {Rey}, {Benneke}, {H{\'e}brard}, {Lovis}, {Demory},
  {Moutou}, \& {Ehrenreich}}]{Bourrier18_55Cnce}
{Bourrier}, V., {Dumusque}, X., {Dorn}, C., {et~al.} 2018, \aap, 619, A1,
  \dodoi{10.1051/0004-6361/201833154}

\bibitem[{{Bower} {et~al.}(2019){Bower}, {Kitzmann}, {Wolf}, {Sanan}, {Dorn},
  \& {Oza}}]{Bower19MagmaOceanPlanets}
{Bower}, D.~J., {Kitzmann}, D., {Wolf}, A.~S., {et~al.} 2019, \aap, 631, A103,
  \dodoi{10.1051/0004-6361/201935710}

\bibitem[{{Brugger} {et~al.}(2017){Brugger}, {Mousis}, {Deleuil}, \&
  {Deschamps}}]{Brugger18_SuperEarth_Interiors}
{Brugger}, B., {Mousis}, O., {Deleuil}, M., \& {Deschamps}, F. 2017, \apj, 850,
  93, \dodoi{10.3847/1538-4357/aa965a}

\bibitem[{{Buchhave} {et~al.}(2016){Buchhave}, {Dressing}, {Dumusque}, {Rice},
  {Vanderburg}, {Mortier}, {Lopez-Morales}, {Lopez}, {Lundkvist}, {Kjeldsen},
  {Affer}, {Bonomo}, {Charbonneau}, {Collier Cameron}, {Cosentino}, {Figueira},
  {Fiorenzano}, {Harutyunyan}, {Haywood}, {Johnson}, {Latham}, {Lovis},
  {Malavolta}, {Mayor}, {Micela}, {Molinari}, {Motalebi}, {Nascimbeni}, {Pepe},
  {Phillips}, {Piotto}, {Pollacco}, {Queloz}, {Sasselov}, {S{\'e}gransan},
  {Sozzetti}, {Udry}, \& {Watson}}]{Buchhave16_Kep20b_MR}
{Buchhave}, L.~A., {Dressing}, C.~D., {Dumusque}, X., {et~al.} 2016, \aj, 152,
  160, \dodoi{10.3847/0004-6256/152/6/160}

\bibitem[{Chau {et~al.}(2018)Chau, Reinhardt, Helled, \& Stadel}]{Chau_2018}
Chau, A., Reinhardt, C., Helled, R., \& Stadel, J. 2018, The Astrophysical
  Journal, 865, 35, \dodoi{10.3847/1538-4357/aad8b0}

\bibitem[{{Cloutier} {et~al.}(2019){Cloutier}, {Astudillo-Defru}, {Bonfils},
  {Jenkins}, {Berdi{\~n}as}, {Ricker}, {Vand erspek}, {Latham}, {Seager},
  {Winn}, {Jenkins}, {Almenara}, {Bouchy}, {Delfosse}, {D{\'\i}az},
  {D{\'\i}az}, {Doyon}, {Figueira}, {Forveille}, {Kurtovic}, {Lovis}, {Mayor},
  {Menou}, {Morgan}, {Morris}, {Muirhead}, {Murgas}, {Pepe}, {Santos},
  {S{\'e}gransan}, {Smith}, {Tenenbaum}, {Torres}, {Udry}, {Vezie}, \&
  {Villasenor}}]{Cloutier19_L9858cd_MR}
{Cloutier}, R., {Astudillo-Defru}, N., {Bonfils}, X., {et~al.} 2019, \aap, 629,
  A111, \dodoi{10.1051/0004-6361/201935957}

\bibitem[{{Cloutier} {et~al.}(2020{\natexlab{a}}){Cloutier}, {Eastman},
  {Rodriguez}, {Astudillo-Defru}, {Bonfils}, {Mortier}, {Watson}, {Stalport},
  {Pinamonti}, {Lienhard}, {Harutyunyan}, {Damasso}, {Latham}, {Collins},
  {Massey}, {Irwin}, {Winters}, {Charbonneau}, {Ziegler}, {Matthews},
  {Crossfield}, {Kreidberg}, {Quinn}, {Ricker}, {Vanderspek}, {Seager}, {Winn},
  {Jenkins}, {Vezie}, {Udry}, {Twicken}, {Tenenbaum}, {Sozzetti},
  {S{\'e}gransan}, {Schlieder}, {Sasselov}, {Santos}, {Rice}, {Rackham},
  {Poretti}, {Piotto}, {Phillips}, {Pepe}, {Molinari}, {Mignon}, {Micela},
  {Melo}, {de Medeiros}, {Mayor}, {Matson}, {Martinez Fiorenzano}, {Mann},
  {Magazz{\'u}}, {Lovis}, {L{\'o}pez-Morales}, {Lopez}, {Lissauer},
  {L{\'e}pine}, {Law}, {Kielkopf}, {Johnson}, {Jensen}, {Howell}, {Gonzales},
  {Ghedina}, {Forveille}, {Figueira}, {Dumusque}, {Dressing}, {Doyon},
  {D{\'\i}az}, {Fabrizio}, {Delfosse}, {Cosentino}, {Conti}, {Collins},
  {Cameron}, {Ciardi}, {Caldwell}, {Burke}, {Buchhave}, {Brice{\~n}o}, {Boyd},
  {Bouchy}, {Beichman}, {Artigau}, \& {Almenara}}]{Cloutier20_LT3780_MR}
{Cloutier}, R., {Eastman}, J.~D., {Rodriguez}, J.~E., {et~al.}
  2020{\natexlab{a}}, \aj, 160, 3, \dodoi{10.3847/1538-3881/ab91c2}

\bibitem[{{Cloutier} {et~al.}(2020{\natexlab{b}}){Cloutier}, {Rodriguez},
  {Irwin}, {Charbonneau}, {Stassun}, {Mortier}, {Latham}, {Isaacson}, {Howard},
  {Udry}, {Wilson}, {Watson}, {Pinamonti}, {Lienhard}, {Giacobbe}, {Guerra},
  {Collins}, {Beiryla}, {Esquerdo}, {Matthews}, {Matson}, {Howell}, {Furlan},
  {Crossfield}, {Winters}, {Nava}, {Ment}, {Lopez}, {Ricker}, {Vanderspek},
  {Seager}, {Jenkins}, {Ting}, {Tenenbaum}, {Sozzetti}, {Sha}, {S{\'e}gransan},
  {Schlieder}, {Sasselov}, {Roy}, {Robertson}, {Rice}, {Poretti}, {Piotto},
  {Phillips}, {Pepper}, {Pepe}, {Molinari}, {Mocnik}, {Micela}, {Mayor},
  {Martinez Fiorenzano}, {Mallia}, {Lubin}, {Lovis}, {L{\'o}pez-Morales},
  {Kosiarek}, {Kielkopf}, {Kane}, {Jensen}, {Isopi}, {Huber}, {Hill},
  {Harutyunyan}, {Gonzales}, {Giacalone}, {Ghedina}, {Ercolino}, {Dumusque},
  {Dressing}, {Damasso}, {Dalba}, {Cosentino}, {Conti}, {Col{\'o}n}, {Collins},
  {Cameron}, {Ciardi}, {Christiansen}, {Chontos}, {Cecconi}, {Caldwell},
  {Burke}, {Buchhave}, {Beichman}, {Behmard}, {Beard}, \& {Akana
  Murphy}}]{Cloutier20_TOI1235b_MR}
{Cloutier}, R., {Rodriguez}, J.~E., {Irwin}, J., {et~al.} 2020{\natexlab{b}},
  \aj, 160, 22, \dodoi{10.3847/1538-3881/ab9534}

\bibitem[{{Crida} {et~al.}(2018){Crida}, {Ligi}, {Dorn}, {Borsa}, \&
  {Lebreton}}]{Crida18a_55Cnce}
{Crida}, A., {Ligi}, R., {Dorn}, C., {Borsa}, F., \& {Lebreton}, Y. 2018,
  Research Notes of the American Astronomical Society, 2, 172,
  \dodoi{10.3847/2515-5172/aae1f6}

\bibitem[{Crida {et~al.}(2018)Crida, Ligi, Dorn, Borsa, \&
  Lebreton}]{Crida_2018}
Crida, A., Ligi, R., Dorn, C., Borsa, F., \& Lebreton, Y. 2018, Research Notes
  of the {AAS}, 2, 172, \dodoi{10.3847/2515-5172/aae1f6}

\bibitem[{{Crida} {et~al.}(2018){Crida}, {Ligi}, {Dorn}, \&
  {Lebreton}}]{Crida18_55Cnce}
{Crida}, A., {Ligi}, R., {Dorn}, C., \& {Lebreton}, Y. 2018, \apj, 860, 122,
  \dodoi{10.3847/1538-4357/aabfe4}

\bibitem[{{Dai} {et~al.}(2019){Dai}, {Masuda}, {Winn}, \&
  {Zeng}}]{Dai_Homogenous_MRs_HotEarths}
{Dai}, F., {Masuda}, K., {Winn}, J.~N., \& {Zeng}, L. 2019, \apj, 883, 79,
  \dodoi{10.3847/1538-4357/ab3a3b}

\bibitem[{{Dai} {et~al.}(2015){Dai}, {Winn}, {Arriagada}, {Butler}, {Crane},
  {Johnson}, {Shectman}, {Teske}, {Thompson}, {Vanderburg}, \&
  {Wittenmyer}}]{Dai15_WASP47e}
{Dai}, F., {Winn}, J.~N., {Arriagada}, P., {et~al.} 2015, \apjl, 813, L9,
  \dodoi{10.1088/2041-8205/813/1/L9}

\bibitem[{{Demory} {et~al.}(2016{\natexlab{a}}){Demory}, {Gillon},
  {Madhusudhan}, \& {Queloz}}]{Demory16_55Cnc}
{Demory}, B.-O., {Gillon}, M., {Madhusudhan}, N., \& {Queloz}, D.
  2016{\natexlab{a}}, \mnras, 455, 2018, \dodoi{10.1093/mnras/stv2239}

\bibitem[{{Demory} {et~al.}(2011){Demory}, {Gillon}, {Deming}, {Valencia},
  {Seager}, {Benneke}, {Lovis}, {Cubillos}, {Harrington}, {Stevenson}, {Mayor},
  {Pepe}, {Queloz}, {S{\'e}gransan}, \& {Udry}}]{Demory11_55Cnce}
{Demory}, B.~O., {Gillon}, M., {Deming}, D., {et~al.} 2011, \aap, 533, A114,
  \dodoi{10.1051/0004-6361/201117178}

\bibitem[{{Demory} {et~al.}(2016{\natexlab{b}}){Demory}, {Gillon}, {de Wit},
  {Madhusudhan}, {Bolmont}, {Heng}, {Kataria}, {Lewis}, {Hu}, {Krick},
  {Stamenkovi{\'c}}, {Benneke}, {Kane}, \& {Queloz}}]{Demory16b_55Cnce}
{Demory}, B.-O., {Gillon}, M., {de Wit}, J., {et~al.} 2016{\natexlab{b}}, \nat,
  532, 207, \dodoi{10.1038/nature17169}

\bibitem[{{Demory} {et~al.}(2016{\natexlab{c}}){Demory}, {Gillon}, {de Wit},
  {Madhusudhan}, {Bolmont}, {Heng}, {Kataria}, {Lewis}, {Hu}, {Krick},
  {Stamenkovi{\'c}}, {Benneke}, {Kane}, \& {Queloz}}]{Demory16_55Cnce}
---. 2016{\natexlab{c}}, \nat, 532, 207, \dodoi{10.1038/nature17169}

\bibitem[{{Deng}(2020)}]{Deng20_Planet-Planet_tidal_stripping}
{Deng}, H. 2020, \apjl, 888, L1, \dodoi{10.3847/2041-8213/ab6084}

\bibitem[{{Dorn} {et~al.}(2019){Dorn}, {Harrison}, {Bonsor}, \&
  {Hands}}]{Dorn_CAI_Planets}
{Dorn}, C., {Harrison}, J.~H.~D., {Bonsor}, A., \& {Hands}, T.~O. 2019, \mnras,
  484, 712, \dodoi{10.1093/mnras/sty3435}

\bibitem[{{Dorn} {et~al.}(2017){Dorn}, {Hinkel}, \& {Venturini}}]{dorn17_bayes}
{Dorn}, C., {Hinkel}, N.~R., \& {Venturini}, J. 2017, \aap, 597, A38,
  \dodoi{10.1051/0004-6361/201628749}

\bibitem[{{Dorn} {et~al.}(2015){Dorn}, {Khan}, {Heng}, {Connolly}, {Alibert},
  {Benz}, \& {Tackley}}]{Dorn15_A&A_bayesian_rocky_interiors}
{Dorn}, C., {Khan}, A., {Heng}, K., {et~al.} 2015, \aap, 577, A83,
  \dodoi{10.1051/0004-6361/201424915}

\bibitem[{{Dumusque} {et~al.}(2014){Dumusque}, {Bonomo}, {Haywood},
  {Malavolta}, {S{\'e}gransan}, {Buchhave}, {Collier Cameron}, {Latham},
  {Molinari}, {Pepe}, {Udry}, {Charbonneau}, {Cosentino}, {Dressing},
  {Figueira}, {Fiorenzano}, {Gettel}, {Harutyunyan}, {Horne}, {Lopez-Morales},
  {Lovis}, {Mayor}, {Micela}, {Motalebi}, {Nascimbeni}, {Phillips}, {Piotto},
  {Pollacco}, {Queloz}, {Rice}, {Sasselov}, {Sozzetti}, {Szentgyorgyi}, \&
  {Watson}}]{Dumusque_Kep10b}
{Dumusque}, X., {Bonomo}, A.~S., {Haywood}, R.~D., {et~al.} 2014, \apj, 789,
  154, \dodoi{10.1088/0004-637X/789/2/154}

\bibitem[{{Dumusque} {et~al.}(2019){Dumusque}, {Turner}, {Dorn}, {Eastman},
  {Allart}, {Adibekyan}, {Sousa}, {Santos}, {Mordasini}, {Bourrier}, {Bouchy},
  {Coffinet}, {Davies}, {D{\'\i}az}, {Fausnaugh}, {Glidden}, {Guerrero},
  {Henze}, {Jenkins}, {Latham}, {Lovis}, {Mayor}, {Pepe}, {Quintana}, {Ricker},
  {Rowden}, {Segransan}, {Mascare{\~n}o}, {Seager}, {Twicken}, {Udry}, {Vand
  erspek}, \& {Winn}}]{HD15337_MR_Dumusque}
{Dumusque}, X., {Turner}, O., {Dorn}, C., {et~al.} 2019, \aap, 627, A43,
  \dodoi{10.1051/0004-6361/201935457}

\bibitem[{Ebel \& Stewart(2019)}]{ebel2019}
Ebel, D., \& Stewart, S. 2019, in Mercury: The view after MESSENGER, ed.
  {Solomon, S. C.}, {Nittler, L. R.}, \& {Anderson, B. J.} (Oxford: Cambridge
  University Press)

\bibitem[{{Ehrenreich} {et~al.}(2012){Ehrenreich}, {Bourrier}, {Bonfils},
  {Lecavelier des Etangs}, {H{\'e}brard}, {Sing}, {Wheatley}, {Vidal-Madjar},
  {Delfosse}, {Udry}, {Forveille}, \& {Moutou}}]{ehrenreich_volatile}
{Ehrenreich}, D., {Bourrier}, V., {Bonfils}, X., {et~al.} 2012, \aap, 547, A18,
  \dodoi{10.1051/0004-6361/201219981}

\bibitem[{Elkins-Tanton \& Seager(2008)}]{Elkins_Tanton_2008}
Elkins-Tanton, L.~T., \& Seager, S. 2008, The Astrophysical Journal, 688, 628,
  \dodoi{10.1086/592316}

\bibitem[{{Endl} {et~al.}(2012){Endl}, {Robertson}, {Cochran}, {MacQueen},
  {Brugamyer}, {Caldwell}, {Wittenmyer}, {Barnes}, \&
  {Gullikson}}]{Endl_55Cnce}
{Endl}, M., {Robertson}, P., {Cochran}, W.~D., {et~al.} 2012, \apj, 759, 19,
  \dodoi{10.1088/0004-637X/759/1/19}

\bibitem[{{Epstein} {et~al.}(2010){Epstein}, {Johnson}, {Dong}, {Udalski},
  {Gould}, \& {Becker}}]{Epstein10}
{Epstein}, C.~R., {Johnson}, J.~A., {Dong}, S., {et~al.} 2010, \apj, 709, 447,
  \dodoi{10.1088/0004-637X/709/1/447}

\bibitem[{{Espinoza} {et~al.}(2019){Espinoza}, {Brahm}, {Henning},
  {Jord{\'a}n}, {Dorn}, {Rojas}, {Sarkis}, {Kossakowski}, {Schlecker},
  {D{\'\i}az}, {Jenkins}, {Aguilera-Gomez}, {Jenkins}, {Twicken}, {Collins},
  {Lissauer}, {Armstrong}, {Adibekyan}, {Barrado}, {Barros}, {Battley},
  {Bayliss}, {Bouchy}, {Bryant}, {Cooke}, {Demangeon}, {Dumusque}, {Figueira},
  {Giles}, {Lillo-Box}, {Lovis}, {Nielsen}, {Pepe}, {Pollacco}, {Santos},
  {Sousa}, {Udry}, {Wheatley}, {Turner}, {Marmier}, {S{\'e}gransan}, {Ricker},
  {Latham}, {Seager}, {Winn}, {Kielkopf}, {Hart}, {Wingham}, {Jensen},
  {He{\l}miniak}, {Tokovinin}, {Brice{\~n}o}, {Ziegler}, {Law}, {Mann},
  {Daylan}, {Doty}, {Guerrero}, {Boyd}, {Crossfield}, {Morris}, {Henze}, \&
  {Chacon}}]{Espinoza_HD213885}
{Espinoza}, N., {Brahm}, R., {Henning}, T., {et~al.} 2019, \mnras, 2769,
  \dodoi{10.1093/mnras/stz3150}

\bibitem[{{Esteves} {et~al.}(2015){Esteves}, {De Mooij}, \&
  {Jayawardhana}}]{Esteves_Kep10b}
{Esteves}, L.~J., {De Mooij}, E. J.~W., \& {Jayawardhana}, R. 2015, \apj, 804,
  150, \dodoi{10.1088/0004-637X/804/2/150}

\bibitem[{{Fogtmann-Schulz} {et~al.}(2014){Fogtmann-Schulz}, {Hinrup}, {Van
  Eylen}, {Christensen-Dalsgaard}, {Kjeldsen}, {Silva Aguirre}, \&
  {Tingley}}]{Fogtmann_kep10b}
{Fogtmann-Schulz}, A., {Hinrup}, B., {Van Eylen}, V., {et~al.} 2014, \apj, 781,
  67, \dodoi{10.1088/0004-637X/781/2/67}

\bibitem[{{Frustagli} {et~al.}(2020){Frustagli}, {Poretti}, {Milbourne},
  {Malavolta}, {Mortier}, {Singh}, {Bonomo}, {Buchhave}, {Zeng}, {Vanderburg},
  {Udry}, {Andreuzzi}, {Collier-Cameron}, {Cosentino}, {Damasso}, {Ghedina},
  {Harutyunyan}, {Haywood}, {Latham}, {L{\'o}pez-Morales}, {Lorenzi}, {Martinez
  Fiorenzano}, {Mayor}, {Micela}, {Molinari}, {Pepe}, {Phillips}, {Rice}, \&
  {Sozzetti}}]{Frustagli20_HD80653b_MR}
{Frustagli}, G., {Poretti}, E., {Milbourne}, T., {et~al.} 2020, \aap, 633,
  A133, \dodoi{10.1051/0004-6361/201936689}

\bibitem[{{Gaia Collaboration} {et~al.}(2016){Gaia Collaboration}, {Prusti},
  {de Bruijne}, {Brown}, {Vallenari}, {Babusiaux}, {Bailer-Jones}, {Bastian},
  {Biermann}, {Evans}, {Eyer}, {Jansen}, {Jordi}, {Klioner}, {Lammers},
  {Lindegren}, {Luri}, {Mignard}, {Milligan}, {Panem}, {Poinsignon},
  {Pourbaix}, {Randich}, {Sarri}, {Sartoretti}, {Siddiqui}, {Soubiran},
  {Valette}, {van Leeuwen}, {Walton}, {Aerts}, {Arenou}, {Cropper}, {Drimmel},
  {H{\o}g}, {Katz}, {Lattanzi}, {O'Mullane}, {Grebel}, {Holland}, {Huc},
  {Passot}, {Bramante}, {Cacciari}, {Casta{\~n}eda}, {Chaoul}, {Cheek}, {De
  Angeli}, {Fabricius}, {Guerra}, {Hern{\'a}ndez}, {Jean-Antoine-Piccolo},
  {Masana}, {Messineo}, {Mowlavi}, {Nienartowicz}, {Ord{\'o}{\~n}ez-Blanco},
  {Panuzzo}, {Portell}, {Richards}, {Riello}, {Seabroke}, {Tanga},
  {Th{\'e}venin}, {Torra}, {Els}, {Gracia-Abril}, {Comoretto},
  {Garcia-Reinaldos}, {Lock}, {Mercier}, {Altmann}, {Andrae}, {Astraatmadja},
  {Bellas-Velidis}, {Benson}, {Berthier}, {Blomme}, {Busso}, {Carry},
  {Cellino}, {Clementini}, {Cowell}, {Creevey}, {Cuypers}, {Davidson}, {De
  Ridder}, {de Torres}, {Delchambre}, {Dell'Oro}, {Ducourant}, {Fr{\'e}mat},
  {Garc{\'\i}a-Torres}, {Gosset}, {Halbwachs}, {Hambly}, {Harrison}, {Hauser},
  {Hestroffer}, {Hodgkin}, {Huckle}, {Hutton}, {Jasniewicz}, {Jordan},
  {Kontizas}, {Korn}, {Lanzafame}, {Manteiga}, {Moitinho}, {Muinonen},
  {Osinde}, {Pancino}, {Pauwels}, {Petit}, {Recio-Blanco}, {Robin}, {Sarro},
  {Siopis}, {Smith}, {Smith}, {Sozzetti}, {Thuillot}, {van Reeven}, {Viala},
  {Abbas}, {Abreu Aramburu}, {Accart}, {Aguado}, {Allan}, {Allasia},
  {Altavilla}, {{\'A}lvarez}, {Alves}, {Anderson}, {Andrei}, {Anglada Varela},
  {Antiche}, {Antoja}, {Ant{\'o}n}, {Arcay}, {Atzei}, {Ayache}, {Bach},
  {Baker}, {Balaguer-N{\'u}{\~n}ez}, {Barache}, {Barata}, {Barbier}, {Barblan},
  {Baroni}, {Barrado y Navascu{\'e}s}, {Barros}, {Barstow}, {Becciani},
  {Bellazzini}, {Bellei}, {Bello Garc{\'\i}a}, {Belokurov}, {Bendjoya},
  {Berihuete}, {Bianchi}, {Bienaym{\'e}}, {Billebaud}, {Blagorodnova},
  {Blanco-Cuaresma}, {Boch}, {Bombrun}, {Borrachero}, {Bouquillon}, {Bourda},
  {Bouy}, {Bragaglia}, {Breddels}, {Brouillet}, {Br{\"u}semeister},
  {Bucciarelli}, {Budnik}, {Burgess}, {Burgon}, {Burlacu}, {Busonero}, {Buzzi},
  {Caffau}, {Cambras}, {Campbell}, {Cancelliere}, {Cantat-Gaudin}, {Carlucci},
  {Carrasco}, {Castellani}, {Charlot}, {Charnas}, {Charvet}, {Chassat},
  {Chiavassa}, {Clotet}, {Cocozza}, {Collins}, {Collins}, {Costigan}, {Crifo},
  {Cross}, {Crosta}, {Crowley}, {Dafonte}, {Damerdji}, {Dapergolas}, {David},
  {David}, {De Cat}, {de Felice}, {de Laverny}, {De Luise}, {De March}, {de
  Martino}, {de Souza}, {Debosscher}, {del Pozo}, {Delbo}, {Delgado},
  {Delgado}, {di Marco}, {Di Matteo}, {Diakite}, {Distefano}, {Dolding}, {Dos
  Anjos}, {Drazinos}, {Dur{\'a}n}, {Dzigan}, {Ecale}, {Edvardsson}, {Enke},
  {Erdmann}, {Escolar}, {Espina}, {Evans}, {Eynard Bontemps}, {Fabre},
  {Fabrizio}, {Faigler}, {Falc{\~a}o}, {Farr{\`a}s Casas}, {Faye}, {Federici},
  {Fedorets}, {Fern{\'a}ndez-Hern{\'a}ndez}, {Fernique}, {Fienga}, {Figueras},
  {Filippi}, {Findeisen}, {Fonti}, {Fouesneau}, {Fraile}, {Fraser}, {Fuchs},
  {Furnell}, {Gai}, {Galleti}, {Galluccio}, {Garabato}, {Garc{\'\i}a-Sedano},
  {Gar{\'e}}, {Garofalo}, {Garralda}, {Gavras}, {Gerssen}, {Geyer}, {Gilmore},
  {Girona}, {Giuffrida}, {Gomes}, {Gonz{\'a}lez-Marcos},
  {Gonz{\'a}lez-N{\'u}{\~n}ez}, {Gonz{\'a}lez-Vidal}, {Granvik}, {Guerrier},
  {Guillout}, {Guiraud}, {G{\'u}rpide}, {Guti{\'e}rrez-S{\'a}nchez}, {Guy},
  {Haigron}, {Hatzidimitriou}, {Haywood}, {Heiter}, {Helmi}, {Hobbs},
  {Hofmann}, {Holl}, {Holland }, {Hunt}, {Hypki}, {Icardi}, {Irwin}, {Jevardat
  de Fombelle}, {Jofr{\'e}}, {Jonker}, {Jorissen}, {Julbe}, {Karampelas},
  {Kochoska}, {Kohley}, {Kolenberg}, {Kontizas}, {Koposov}, {Kordopatis},
  {Koubsky}, {Kowalczyk}, {Krone-Martins}, {Kudryashova}, {Kull}, {Bachchan},
  {Lacoste-Seris}, {Lanza}, {Lavigne}, {Le Poncin-Lafitte}, {Lebreton},
  {Lebzelter}, {Leccia}, {Leclerc}, {Lecoeur-Taibi}, {Lemaitre}, {Lenhardt},
  {Leroux}, {Liao}, {Licata}, {Lindstr{\o}m}, {Lister}, {Livanou}, {Lobel},
  {L{\"o}ffler}, {L{\'o}pez}, {Lopez-Lozano}, {Lorenz}, {Loureiro},
  {MacDonald}, {Magalh{\~a}es Fernandes}, {Managau}, {Mann}, {Mantelet},
  {Marchal}, {Marchant}, {Marconi}, {Marie}, {Marinoni}, {Marrese},
  {Marschalk{\'o}}, {Marshall}, {Mart{\'\i}n-Fleitas}, {Martino}, {Mary},
  {Matijevi{\v{c}}}, {Mazeh}, {McMillan}, {Messina}, {Mestre}, {Michalik},
  {Millar}, {Miranda}, {Molina}, {Molinaro}, {Molinaro}, {Moln{\'a}r},
  {Moniez}, {Montegriffo}, {Monteiro}, {Mor}, {Mora}, {Morbidelli}, {Morel},
  {Morgenthaler}, {Morley}, {Morris}, {Mulone}, {Muraveva}, {Musella},
  {Narbonne}, {Nelemans}, {Nicastro}, {Noval}, {Ord{\'e}novic},
  {Ordieres-Mer{\'e}}, {Osborne}, {Pagani}, {Pagano}, {Pailler}, {Palacin},
  {Palaversa}, {Parsons}, {Paulsen}, {Pecoraro}, {Pedrosa}, {Pentik{\"a}inen},
  {Pereira}, {Pichon}, {Piersimoni}, {Pineau}, {Plachy}, {Plum}, {Poujoulet},
  {Pr{\v{s}}a}, {Pulone}, {Ragaini}, {Rago}, {Rambaux}, {Ramos-Lerate},
  {Ranalli}, {Rauw}, {Read}, {Regibo}, {Renk}, {Reyl{\'e}}, {Ribeiro},
  {Rimoldini}, {Ripepi}, {Riva}, {Rixon}, {Roelens}, {Romero-G{\'o}mez},
  {Rowell}, {Royer}, {Rudolph}, {Ruiz-Dern}, {Sadowski}, {Sagrist{\`a}
  Sell{\'e}s}, {Sahlmann}, {Salgado}, {Salguero}, {Sarasso}, {Savietto},
  {Schnorhk}, {Schultheis}, {Sciacca}, {Segol}, {Segovia}, {Segransan},
  {Serpell}, {Shih}, {Smareglia}, {Smart}, {Smith}, {Solano}, {Solitro},
  {Sordo}, {Soria Nieto}, {Souchay}, {Spagna}, {Spoto}, {Stampa}, {Steele},
  {Steidelm{\"u}ller}, {Stephenson}, {Stoev}, {Suess}, {S{\"u}veges}, {Surdej},
  {Szabados}, {Szegedi-Elek}, {Tapiador}, {Taris}, {Tauran}, {Taylor},
  {Teixeira}, {Terrett}, {Tingley}, {Trager}, {Turon}, {Ulla}, {Utrilla},
  {Valentini}, {van Elteren}, {Van Hemelryck}, {van Leeuwen}, {Varadi},
  {Vecchiato}, {Veljanoski}, {Via}, {Vicente}, {Vogt}, {Voss}, {Votruba},
  {Voutsinas}, {Walmsley}, {Weiler}, {Weingrill}, {Werner}, {Wevers},
  {Whitehead}, {Wyrzykowski}, {Yoldas}, {{\v{Z}}erjal}, {Zucker}, {Zurbach},
  {Zwitter}, {Alecu}, {Allen}, {Allende Prieto}, {Amorim},
  {Anglada-Escud{\'e}}, {Arsenijevic}, {Azaz}, {Balm}, {Beck}, {Bernstein},
  {Bigot}, {Bijaoui}, {Blasco}, {Bonfigli}, {Bono}, {Boudreault}, {Bressan},
  {Brown}, {Brunet}, {Bunclark}, {Buonanno}, {Butkevich}, {Carret}, {Carrion},
  {Chemin}, {Ch{\'e}reau}, {Corcione}, {Darmigny}, {de Boer}, {de Teodoro}, {de
  Zeeuw}, {Delle Luche}, {Domingues}, {Dubath}, {Fodor}, {Fr{\'e}zouls},
  {Fries}, {Fustes}, {Fyfe}, {Gallardo}, {Gallegos}, {Gardiol}, {Gebran},
  {Gomboc}, {G{\'o}mez}, {Grux}, {Gueguen}, {Heyrovsky}, {Hoar}, {Iannicola},
  {Isasi Parache}, {Janotto}, {Joliet}, {Jonckheere}, {Keil}, {Kim},
  {Klagyivik}, {Klar}, {Knude}, {Kochukhov}, {Kolka}, {Kos}, {Kutka}, {Lainey},
  {LeBouquin}, {Liu}, {Loreggia}, {Makarov}, {Marseille}, {Martayan},
  {Martinez-Rubi}, {Massart}, {Meynadier}, {Mignot}, {Munari}, {Nguyen},
  {Nordlander}, {Ocvirk}, {O'Flaherty}, {Olias Sanz}, {Ortiz}, {Osorio},
  {Oszkiewicz}, {Ouzounis}, {Palmer}, {Park}, {Pasquato}, {Peltzer}, {Peralta},
  {P{\'e}turaud}, {Pieniluoma}, {Pigozzi}, {Poels}, {Prat}, {Prod'homme},
  {Raison}, {Rebordao}, {Risquez}, {Rocca-Volmerange}, {Rosen}, {Ruiz-Fuertes},
  {Russo}, {Sembay}, {Serraller Vizcaino}, {Short}, {Siebert}, {Silva},
  {Sinachopoulos}, {Slezak}, {Soffel}, {Sosnowska}, {Strai{\v{z}}ys}, {ter
  Linden}, {Terrell}, {Theil}, {Tiede}, {Troisi}, {Tsalmantza}, {Tur},
  {Vaccari}, {Vachier}, {Valles}, {Van Hamme}, {Veltz}, {Virtanen}, {Wallut},
  {Wichmann}, {Wilkinson}, {Ziaeepour}, \& {Zschocke}}]{gaia}
{Gaia Collaboration}, {Prusti}, T., {de Bruijne}, J.~H.~J., {et~al.} 2016,
  \aap, 595, A1, \dodoi{10.1051/0004-6361/201629272}

\bibitem[{{Gandolfi} {et~al.}(2019){Gandolfi}, {Fossati}, {Livingston},
  {Stassun}, {Grziwa}, {Barrag{\'a}n}, {Fridlund}, {Kubyshkina}, {Persson},
  {Dai}, {Lam}, {Albrecht}, {Batalha}, {Beck}, {Justesen}, {Cabrera},
  {Cartwright}, {Cochran}, {Csizmadia}, {Davies}, {Deeg}, {Eigm{\"u}ller},
  {Endl}, {Erikson}, {Esposito}, {Garc{\'\i}a}, {Goeke}, {Gonz{\'a}lez-Cuesta},
  {Guenther}, {Hatzes}, {Hidalgo}, {Hirano}, {Hjorth}, {Kabath}, {Knudstrup},
  {Korth}, {Li}, {Luque}, {Mathur}, {Monta{\~n}es Rodr{\'\i}guez}, {Narita},
  {Nespral}, {Niraula}, {Nowak}, {Palle}, {P{\"a}tzold}, {Prieto-Arranz},
  {Rauer}, {Redfield}, {Ribas}, {Skarka}, {Smith}, {Rowden}, {Torres}, {Van
  Eylen}, \& {Vezie}}]{Gandolfi19_HD15337b}
{Gandolfi}, D., {Fossati}, L., {Livingston}, J.~H., {et~al.} 2019, \apjl, 876,
  L24, \dodoi{10.3847/2041-8213/ab17d9}

\bibitem[{{Gautier} {et~al.}(2012){Gautier}, {Charbonneau}, {Rowe}, {Marcy},
  {Isaacson}, {Torres}, {Fressin}, {Rogers}, {D{\'e}sert}, {Buchhave},
  {Latham}, {Quinn}, {Ciardi}, {Fabrycky}, {Ford}, {Gilliland}, {Walkowicz},
  {Bryson}, {Cochran}, {Endl}, {Fischer}, {Howell}, {Horch}, {Barclay},
  {Batalha}, {Borucki}, {Christiansen}, {Geary}, {Henze}, {Holman}, {Ibrahim},
  {Jenkins}, {Kinemuchi}, {Koch}, {Lissauer}, {Sanderfer}, {Sasselov},
  {Seager}, {Silverio}, {Smith}, {Still}, {Stumpe}, {Tenenbaum}, \& {Van
  Cleve}}]{Gautier12_Kep20b}
{Gautier}, Thomas~N., I., {Charbonneau}, D., {Rowe}, J.~F., {et~al.} 2012,
  \apj, 749, 15, \dodoi{10.1088/0004-637X/749/1/15}

\bibitem[{{Gillon} {et~al.}(2017){Gillon}, {Demory}, {Van Grootel}, {Motalebi},
  {Lovis}, {Cameron}, {Charbonneau}, {Latham}, {Molinari}, {Pepe},
  {S{\'e}gransan}, {Sasselov}, {Udry}, {Mayor}, {Micela}, {Piotto}, \&
  {Sozzetti}}]{HD213194bc_Gillon}
{Gillon}, M., {Demory}, B.-O., {Van Grootel}, V., {et~al.} 2017, Nature
  Astronomy, 1, 0056, \dodoi{10.1038/s41550-017-0056}

\bibitem[{{Helffrich}(2017)}]{Mars_core_and_LEs}
{Helffrich}, G. 2017, Progress in Earth and Planetary Science, 4, 24,
  \dodoi{10.1186/s40645-017-0139-4}

\bibitem[{{Hellier} {et~al.}(2012){Hellier}, {Anderson}, {Collier Cameron},
  {Doyle}, {Fumel}, {Gillon}, {Jehin}, {Lendl}, {Maxted}, {Pepe}, {Pollacco},
  {Queloz}, {S{\'e}gransan}, {Smalley}, {Smith}, {Southworth}, {Triaud},
  {Udry}, \& {West}}]{Hellier12_WASP47}
{Hellier}, C., {Anderson}, D.~R., {Collier Cameron}, A., {et~al.} 2012, \mnras,
  426, 739, \dodoi{10.1111/j.1365-2966.2012.21780.x}

\bibitem[{{Hinkel} {et~al.}(2014){Hinkel}, {Timmes}, {Young}, {Pagano}, \&
  {Turnbull}}]{Hinkel14_Hypatia}
{Hinkel}, N.~R., {Timmes}, F.~X., {Young}, P.~A., {Pagano}, M.~D., \&
  {Turnbull}, M.~C. 2014, \aj, 148, 54, \dodoi{10.1088/0004-6256/148/3/54}

\bibitem[{Jia \& Spruit(2016)}]{Shi16_mass_transfer}
Jia, S., \& Spruit, H.~C. 2016, Monthly Notices of the Royal Astronomical
  Society, 465, 149, \dodoi{10.1093/mnras/stw1693}

\bibitem[{{Jin} \& {Mordasini}(2018)}]{Jin_Mordasini18_Photoevap}
{Jin}, S., \& {Mordasini}, C. 2018, \apj, 853, 163,
  \dodoi{10.3847/1538-4357/aa9f1e}

\bibitem[{{Jindal} {et~al.}(2020){Jindal}, {de Mooij}, {Jayawardhana},
  {Deibert}, {Brogi}, {Rustamkulov}, {Fortney}, {Hood}, \&
  {Morley}}]{Jindal2020}
{Jindal}, A., {de Mooij}, E. J.~W., {Jayawardhana}, R., {et~al.} 2020, \aj,
  160, 101, \dodoi{10.3847/1538-3881/aba1eb}

\bibitem[{{Jontof-Hutter} {et~al.}(2016){Jontof-Hutter}, {Ford}, {Rowe},
  {Lissauer}, {Fabrycky}, {Van Laerhoven}, {Agol}, {Deck}, {Holczer}, \&
  {Mazeh}}]{JontofHutter16_Kep60b__kep105c_MR}
{Jontof-Hutter}, D., {Ford}, E.~B., {Rowe}, J.~F., {et~al.} 2016, \apj, 820,
  39, \dodoi{10.3847/0004-637X/820/1/39}

\bibitem[{{Koll} {et~al.}(2019){Koll}, {Malik}, {Mansfield}, {Kempton}, {Kite},
  {Abbot}, \& {Bean}}]{Koll19}
{Koll}, D. D.~B., {Malik}, M., {Mansfield}, M., {et~al.} 2019, \apj, 886, 140,
  \dodoi{10.3847/1538-4357/ab4c91}

\bibitem[{{Kosiarek} {et~al.}(2019){Kosiarek}, {Blunt}, {L{\'o}pez-Morales},
  {Crossfield}, {Sinukoff}, {Petigura}, {Gonzales}, {Poretti}, {Malavolta},
  {Howard}, {Isaacson}, {Haywood}, {Ciardi}, {Bristow}, {Collier Cameron},
  {Charbonneau}, {Dressing}, {Figueira}, {Fulton}, {Hardee}, {Hirsch},
  {Latham}, {Mortier}, {Nava}, {Schlieder}, {Vanderburg}, {Weiss}, {Bonomo},
  {Bouchy}, {Buchhave}, {Coffinet}, {Damasso}, {Dumusque}, {Lovis}, {Mayor},
  {Micela}, {Molinari}, {Pepe}, {Phillips}, {Piotto}, {Rice}, {Sasselov},
  {S{\'e}gransan}, {Sozzetti}, {Udry}, \& {Watson}}]{Kosiarek19_K2291b_MR}
{Kosiarek}, M.~R., {Blunt}, S., {L{\'o}pez-Morales}, M., {et~al.} 2019, \aj,
  157, 116, \dodoi{10.3847/1538-3881/aafe83}

\bibitem[{{Kuchner}(2004)}]{Kuch04}
{Kuchner}, M.~J. 2004, \apj, 612, 1147, \dodoi{10.1086/422577}

\bibitem[{{Lam} {et~al.}(2018){Lam}, {Santerne}, {Sousa}, {Vigan}, {Armstrong},
  {Barros}, {Brugger}, {Adibekyan}, {Almenara}, {Delgado Mena}, {Dumusque},
  {Barrado}, {Bayliss}, {Bonomo}, {Bouchy}, {Brown}, {Ciardi}, {Deleuil},
  {Demangeon}, {Faedi}, {Foxell}, {Jackman}, {King}, {Kirk}, {Ligi},
  {Lillo-Box}, {Lopez}, {Lovis}, {Louden}, {Nielsen}, {McCormac}, {Mousis},
  {Osborn}, {Pollacco}, {Santos}, {Udry}, \& {Wheatley}}]{Lam18_k2265b}
{Lam}, K.~W.~F., {Santerne}, A., {Sousa}, S.~G., {et~al.} 2018, \aap, 620, A77,
  \dodoi{10.1051/0004-6361/201834073}

\bibitem[{{Lee} {et~al.}(2004){Lee}, {O'Neill}, {Panero}, {Shim}, {Benedetti},
  \& {Jeanloz}}]{KKMLee04}
{Lee}, K. K.~M., {O'Neill}, B., {Panero}, W.~R., {et~al.} 2004, Earth and
  Planetary Science Letters, 223, 381, \dodoi{10.1016/j.epsl.2004.04.033}

\bibitem[{{Lehmann}(1936)}]{Lehmann_Earths_Core}
{Lehmann}, I. 1936, Bureau Central Séismologique International Strasbourg:
  Publications du Bureau Central Scientifiques, 87

\bibitem[{Leinhardt \& Stewart(2011)}]{Leinhardt_2011}
Leinhardt, Z.~M., \& Stewart, S.~T. 2011, The Astrophysical Journal, 745, 79,
  \dodoi{10.1088/0004-637x/745/1/79}

\bibitem[{Lewis(1972)}]{LEWIS1972286}
Lewis, J.~S. 1972, Earth and Planetary Science Letters, 15, 286 ,
  \dodoi{https://doi.org/10.1016/0012-821X(72)90174-4}

\bibitem[{{Ligi} {et~al.}(2019){Ligi}, {Dorn}, {Crida}, {Lebreton}, {Creevey},
  {Borsa}, {Mourard}, {Nardetto}, {Tallon-Bosc}, {Morand}, \&
  {Poretti}}]{Ligi19_HD219134}
{Ligi}, R., {Dorn}, C., {Crida}, A., {et~al.} 2019, \aap, 631, A92,
  \dodoi{10.1051/0004-6361/201936259}

\bibitem[{{Liu} {et~al.}(2019){Liu}, {Asplund}, {Yong}, {Feltzing}, {Dotter},
  {Mel{\'e}ndez}, \& {Ram{\'\i}rez}}]{Liu19}
{Liu}, F., {Asplund}, M., {Yong}, D., {et~al.} 2019, \aap, 627, A117,
  \dodoi{10.1051/0004-6361/201935306}

\bibitem[{{Liu} {et~al.}(2016){Liu}, {Yong}, {Asplund}, {Ram{\'\i}rez},
  {Mel{\'e}ndez}, {Gustafsson}, {Howes}, {Roederer}, {Lambert}, \&
  {Bensby}}]{Liu16_kep10_abunds}
{Liu}, F., {Yong}, D., {Asplund}, M., {et~al.} 2016, \mnras, 456, 2636,
  \dodoi{10.1093/mnras/stv2821}

\bibitem[{{Lodders}(2003)}]{Lodders03}
{Lodders}, K. 2003, \apj, 591, 1220, \dodoi{10.1086/375492}

\bibitem[{{Lodders} {et~al.}(2009){Lodders}, {Palme}, \& {Gail}}]{Lodders09}
{Lodders}, K., {Palme}, H., \& {Gail}, H.~P. 2009, Landolt B\&ouml;rnstein, 4B,
  712, \dodoi{10.1007/978-3-540-88055-4_34}

\bibitem[{{Luque} {et~al.}(2019){Luque}, {Pall{\'e}}, {Kossakowski},
  {Dreizler}, {Kemmer}, {Espinoza}, {Burt}, {Anglada-Escud{\'e}}, {B{\'e}jar},
  {Caballero}, {Collins}, {Collins}, {Cort{\'e}s-Contreras},
  {D{\'\i}ez-Alonso}, {Feng}, {Hatzes}, {Hellier}, {Henning}, {Jeffers},
  {Kaltenegger}, {K{\"u}rster}, {Madden}, {Molaverdikhani}, {Montes}, {Narita},
  {Nowak}, {Ofir}, {Oshagh}, {Parviainen}, {Quirrenbach}, {Reffert}, {Reiners},
  {Rodr{\'\i}guez-L{\'o}pez}, {Schlecker}, {Stock}, {Trifonov}, {Winn},
  {Zapatero Osorio}, {Zechmeister}, {Amado}, {Anderson}, {Batalha}, {Bauer},
  {Bluhm}, {Burke}, {Butler}, {Caldwell}, {Chen}, {Crane}, {Dragomir},
  {Dressing}, {Dynes}, {Jenkins}, {Kaminski}, {Klahr}, {Kotani}, {Lafarga},
  {Latham}, {Lewin}, {McDermott}, {Monta{\~n}{\'e}s-Rodr{\'\i}guez}, {Morales},
  {Murgas}, {Nagel}, {Pedraz}, {Ribas}, {Ricker}, {Rowden}, {Seager},
  {Shectman}, {Tamura}, {Teske}, {Twicken}, {Vanderspeck}, {Wang}, \&
  {Wohler}}]{Luque18_GJ357b_MR}
{Luque}, R., {Pall{\'e}}, E., {Kossakowski}, D., {et~al.} 2019, \aap, 628, A39,
  \dodoi{10.1051/0004-6361/201935801}

\bibitem[{{MacDonald} {et~al.}(2016){MacDonald}, {Ragozzine}, {Fabrycky},
  {Ford}, {Holman}, {Isaacson}, {Lissauer}, {Lopez}, {Mazeh}, {Rogers}, {Rowe},
  {Steffen}, \& {Torres}}]{MacDonald16_Kep80d_MR}
{MacDonald}, M.~G., {Ragozzine}, D., {Fabrycky}, D.~C., {et~al.} 2016, \aj,
  152, 105, \dodoi{10.3847/0004-6256/152/4/105}

\bibitem[{{Malavolta} {et~al.}(2018){Malavolta}, {Mayo}, {Louden}, {Rajpaul},
  {Bonomo}, {Buchhave}, {Kreidberg}, {Kristiansen}, {Lopez-Morales}, {Mortier},
  {Vand erburg}, {Coffinet}, {Ehrenreich}, {Lovis}, {Bouchy}, {Charbonneau},
  {Ciardi}, {Collier Cameron}, {Cosentino}, {Crossfield}, {Damasso},
  {Dressing}, {Dumusque}, {Everett}, {Figueira}, {Fiorenzano}, {Gonzales},
  {Haywood}, {Harutyunyan}, {Hirsch}, {Howell}, {Johnson}, {Latham}, {Lopez},
  {Mayor}, {Micela}, {Molinari}, {Nascimbeni}, {Pepe}, {Phillips}, {Piotto},
  {Rice}, {Sasselov}, {S{\'e}gransan}, {Sozzetti}, {Udry}, \&
  {Watson}}]{Malavolta18_k2141bMR}
{Malavolta}, L., {Mayo}, A.~W., {Louden}, T., {et~al.} 2018, \aj, 155, 107,
  \dodoi{10.3847/1538-3881/aaa5b5}

\bibitem[{{Marcus} {et~al.}(2010){Marcus}, {Sasselov}, {Hernquist}, \&
  {Stewart}}]{Marcus10_MinRad_SEs}
{Marcus}, R.~A., {Sasselov}, D., {Hernquist}, L., \& {Stewart}, S.~T. 2010,
  \apjl, 712, L73, \dodoi{10.1088/2041-8205/712/1/L73}

\bibitem[{{Marcy} {et~al.}(2014){Marcy}, {Isaacson}, {Howard}, {Rowe},
  {Jenkins}, {Bryson}, {Latham}, {Howell}, {Gautier}, {Batalha}, {Rogers},
  {Ciardi}, {Fischer}, {Gilliland}, {Kjeldsen}, {Christensen-Dalsgaard},
  {Huber}, {Chaplin}, {Basu}, {Buchhave}, {Quinn}, {Borucki}, {Koch}, {Hunter},
  {Caldwell}, {Van Cleve}, {Kolbl}, {Weiss}, {Petigura}, {Seager}, {Morton},
  {Johnson}, {Ballard}, {Burke}, {Cochran}, {Endl}, {MacQueen}, {Everett},
  {Lissauer}, {Ford}, {Torres}, {Fressin}, {Brown}, {Steffen}, {Charbonneau},
  {Basri}, {Sasselov}, {Winn}, {Sanchis-Ojeda}, {Christiansen}, {Adams},
  {Henze}, {Dupree}, {Fabrycky}, {Fortney}, {Tarter}, {Holman}, {Tenenbaum},
  {Shporer}, {Lucas}, {Welsh}, {Orosz}, {Bedding}, {Campante}, {Davies},
  {Elsworth}, {Handberg}, {Hekker}, {Karoff}, {Kawaler}, {Lund}, {Lundkvist},
  {Metcalfe}, {Miglio}, {Silva Aguirre}, {Stello}, {White}, {Boss}, {Devore},
  {Gould}, {Prsa}, {Agol}, {Barclay}, {Coughlin}, {Brugamyer}, {Mullally},
  {Quintana}, {Still}, {Thompson}, {Morrison}, {Twicken}, {D{\'e}sert},
  {Carter}, {Crepp}, {H{\'e}brard}, {Santerne}, {Moutou}, {Sobeck}, {Hudgins},
  {Haas}, {Robertson}, {Lillo-Box}, \& {Barrado}}]{Marcy14_kep406b_MR}
{Marcy}, G.~W., {Isaacson}, H., {Howard}, A.~W., {et~al.} 2014, \apjs, 210, 20,
  \dodoi{10.1088/0067-0049/210/2/20}

\bibitem[{{McDonough}(2003)}]{McDonough_composition_of_earths_core}
{McDonough}, W.~F. 2003, Treatise on Geochemistry, 2, 568,
  \dodoi{10.1016/B0-08-043751-6/02015-6}

\bibitem[{McDonough(2017)}]{McDonough2017}
McDonough, W.~F. 2017, Earth's Core, ed. W.~M. White (Cham: Springer
  International Publishing), 1--13, \dodoi{10.1007/978-3-319-39193-9_258-1}

\bibitem[{{Morgan} \& {Anders}(1980)}]{Morgan1980}
{Morgan}, J.~W., \& {Anders}, E. 1980, Proceedings of the National Academy of
  Science, 77, 6973, \dodoi{10.1073/pnas.77.12.6973}

\bibitem[{{Morley} {et~al.}(2017){Morley}, {Kreidberg}, {Rustamkulov},
  {Robinson}, \& {Fortney}}]{Morley_JWST_atmochar_17}
{Morley}, C.~V., {Kreidberg}, L., {Rustamkulov}, Z., {Robinson}, T., \&
  {Fortney}, J.~J. 2017, \apj, 850, 121, \dodoi{10.3847/1538-4357/aa927b}

\bibitem[{{Motalebi} {et~al.}(2015){Motalebi}, {Udry}, {Gillon}, {Lovis},
  {S{\'e}gransan}, {Buchhave}, {Demory}, {Malavolta}, {Dressing}, {Sasselov},
  {Rice}, {Charbonneau}, {Collier Cameron}, {Latham}, {Molinari}, {Pepe},
  {Affer}, {Bonomo}, {Cosentino}, {Dumusque}, {Figueira}, {Fiorenzano},
  {Gettel}, {Harutyunyan}, {Haywood}, {Johnson}, {Lopez}, {Lopez-Morales},
  {Mayor}, {Micela}, {Mortier}, {Nascimbeni}, {Philips}, {Piotto}, {Pollacco},
  {Queloz}, {Sozzetti}, {Vand erburg}, \& {Watson}}]{HD219134b_Motalebi_MR}
{Motalebi}, F., {Udry}, S., {Gillon}, M., {et~al.} 2015, \aap, 584, A72,
  \dodoi{10.1051/0004-6361/201526822}

\bibitem[{Nittler {et~al.}(2019)Nittler, Chabot, Grove, \& Peplowski}]{nittler}
Nittler, L., Chabot, N., Grove, T., \& Peplowski, P. 2019, in Mercury: The view
  after MESSENGER, ed. N.~{Solomon, S. C.} \& {Anderson, B. J.} (Oxford:
  Cambridge University Press)

\bibitem[{{Otegi} {et~al.}(2020){Otegi}, {Dorn}, {Helled}, {Bouchy},
  {Haldemann}, \& {Alibert}}]{Otegi20_preprint}
{Otegi}, J.~F., {Dorn}, C., {Helled}, R., {et~al.} 2020, arXiv e-prints,
  arXiv:2006.12353.
\newblock \doarXiv{2006.12353}

\bibitem[{{Persson} {et~al.}(2018){Persson}, {Fridlund}, {Barrag{\'a}n}, {Dai},
  {Gandolfi}, {Hatzes}, {Hirano}, {Grziwa}, {Korth}, {Prieto-Arranz},
  {Fossati}, {Van Eylen}, {Justesen}, {Livingston}, {Kubyshkina}, {Deeg},
  {Guenther}, {Nowak}, {Cabrera}, {Eigm{\"u}ller}, {Csizmadia}, {Smith},
  {Erikson}, {Albrecht}, {Sobrino}, {Cochran}, {Endl}, {Esposito}, {Fukui},
  {Heeren}, {Hidalgo}, {Hjorth}, {Kuzuhara}, {Narita}, {Nespral}, {Palle},
  {P{\"a}tzold}, {Rauer}, {Rodler}, \& {Winn}}]{Persson18_K2216b_MR}
{Persson}, C.~M., {Fridlund}, M., {Barrag{\'a}n}, O., {et~al.} 2018, \aap, 618,
  A33, \dodoi{10.1051/0004-6361/201832867}

\bibitem[{Plotnykov \& Valencia(2020)}]{Plotnykov20}
Plotnykov, M., \& Valencia, D. 2020, Monthly Notices of the Royal Astronomical
  Society, 499, 932, \dodoi{10.1093/mnras/staa2615}

\bibitem[{{Putirka} \& {Rarick}(2019)}]{Putirka_Rarick_AmM19}
{Putirka}, K.~D., \& {Rarick}, J.~C. 2019, American Mineralogist, 104, 817,
  \dodoi{10.2138/am-2019-6787}

\bibitem[{{Rice} {et~al.}(2019){Rice}, {Malavolta}, {Mayo}, {Mortier},
  {Buchhave}, {Affer}, {Vanderburg}, {Lopez-Morales}, {Poretti}, {Zeng},
  {Collier Cameron}, {Damasso}, {Coffinet}, {Latham}, {Bonomo}, {Bouchy},
  {Charbonneau}, {Dumusque}, {Figueira}, {Martinez Fiorenzano}, {Haywood},
  {Johnson}, {Lopez}, {Lovis}, {Mayor}, {Micela}, {Molinari}, {Nascimbeni},
  {Nava}, {Pepe}, {Phillips}, {Piotto}, {Sasselov}, {S{\'e}gransan},
  {Sozzetti}, {Udry}, \& {Watson}}]{Rice19_GJ9827b_MR}
{Rice}, K., {Malavolta}, L., {Mayo}, A., {et~al.} 2019, \mnras, 484, 3731,
  \dodoi{10.1093/mnras/stz130}

\bibitem[{{Ricker} {et~al.}(2015){Ricker}, {Winn}, {Vanderspek}, {Latham},
  {Bakos}, {Bean}, {Berta-Thompson}, {Brown}, {Buchhave}, {Butler}, {Butler},
  {Chaplin}, {Charbonneau}, {Christensen-Dalsgaard}, {Clampin}, {Deming},
  {Doty}, {De Lee}, {Dressing}, {Dunham}, {Endl}, {Fressin}, {Ge}, {Henning},
  {Holman}, {Howard}, {Ida}, {Jenkins}, {Jernigan}, {Johnson}, {Kaltenegger},
  {Kawai}, {Kjeldsen}, {Laughlin}, {Levine}, {Lin}, {Lissauer}, {MacQueen},
  {Marcy}, {McCullough}, {Morton}, {Narita}, {Paegert}, {Palle}, {Pepe},
  {Pepper}, {Quirrenbach}, {Rinehart}, {Sasselov}, {Sato}, {Seager},
  {Sozzetti}, {Stassun}, {Sullivan}, {Szentgyorgyi}, {Torres}, {Udry}, \&
  {Villasenor}}]{tess}
{Ricker}, G.~R., {Winn}, J.~N., {Vanderspek}, R., {et~al.} 2015, Journal of
  Astronomical Telescopes, Instruments, and Systems, 1, 014003,
  \dodoi{10.1117/1.JATIS.1.1.014003}

\bibitem[{{Rogers} \& {Seager}(2010)}]{Rogers10_FeSiO3}
{Rogers}, L.~A., \& {Seager}, S. 2010, \apj, 712, 974,
  \dodoi{10.1088/0004-637X/712/2/974}

\bibitem[{{Santerne} {et~al.}(2018){Santerne}, {Brugger}, {Armstrong},
  {Adibekyan}, {Lillo-Box}, {Gosselin}, {Aguichine}, {Almenara}, {Barrado},
  {Barros}, {Bayliss}, {Boisse}, {Bonomo}, {Bouchy}, {Brown}, {Deleuil},
  {Delgado Mena}, {Demangeon}, {D{\'\i}az}, {Doyle}, {Dumusque}, {Faedi},
  {Faria}, {Figueira}, {Foxell}, {Giles}, {H{\'e}brard}, {Hojjatpanah},
  {Hobson}, {Jackman}, {King}, {Kirk}, {Lam}, {Ligi}, {Lovis}, {Louden},
  {McCormac}, {Mousis}, {Neal}, {Osborn}, {Pepe}, {Pollacco}, {Santos},
  {Sousa}, {Udry}, \& {Vigan}}]{Santerne18_K2229b}
{Santerne}, A., {Brugger}, B., {Armstrong}, D.~J., {et~al.} 2018, Nature
  Astronomy, 2, 393, \dodoi{10.1038/s41550-018-0420-5}

\bibitem[{{Schuler} {et~al.}(2015){Schuler}, {Vaz}, {Katime Santrich}, {Cunha},
  {Smith}, {King}, {Teske}, {Ghezzi}, {Howell}, \&
  {Isaacson}}]{Kep20_abundances}
{Schuler}, S.~C., {Vaz}, Z.~A., {Katime Santrich}, O. o.~J., {et~al.} 2015,
  \apj, 815, 5, \dodoi{10.1088/0004-637X/815/1/5}

\bibitem[{{Scora} {et~al.}(2020){Scora}, {Valencia}, {Morbidelli}, \&
  {Jacobson}}]{Scora20}
{Scora}, J., {Valencia}, D., {Morbidelli}, A., \& {Jacobson}, S. 2020, \mnras,
  493, 4910, \dodoi{10.1093/mnras/staa568}

\bibitem[{{Smith} {et~al.}(2012){Smith}, {Zuber}, {Phillips}, {Solomon},
  {Hauck}, {Lemoine}, {Mazarico}, {Neumann}, {Peale}, {Margot}, {Johnson},
  {Torrence}, {Perry}, {Rowlands}, {Goossens}, {Head}, \&
  {Taylor}}]{Merc_evidence_for_liq_outer_core}
{Smith}, D.~E., {Zuber}, M.~T., {Phillips}, R.~J., {et~al.} 2012, Science, 336,
  214, \dodoi{10.1126/science.1218809}

\bibitem[{{Smith} {et~al.}(2018){Smith}, {Fratanduono}, {Braun}, {Duffy},
  {Wicks}, {Celliers}, {Ali}, {Fernandez-Pa{\~n}ella}, {Kraus}, {Swift},
  {Collins}, \& {Eggert}}]{Smith18_Fe_EoS}
{Smith}, R.~F., {Fratanduono}, D.~E., {Braun}, D.~G., {et~al.} 2018, Nature
  Astronomy, 2, 452, \dodoi{10.1038/s41550-018-0437-9}

\bibitem[{{Stevens} {et~al.}(2018){Stevens}, {Gaudi}, \&
  {Stassun}}]{Stevens18_MR_precision}
{Stevens}, D.~J., {Gaudi}, B.~S., \& {Stassun}, K.~G. 2018, \apj, 862, 53,
  \dodoi{10.3847/1538-4357/aaccf5}

\bibitem[{Stixrude \& Lithgow-Bertelloni(2005)}]{Stixrude05_MgSiO3_EoS}
Stixrude, L., \& Lithgow-Bertelloni, C. 2005, Geophysical Journal
  International, 162, 610, \dodoi{10.1111/j.1365-246X.2005.02642.x}

\bibitem[{{Swain} {et~al.}(2019){Swain}, {Estrela}, {Sotin}, {Roudier}, \&
  {Zellem}}]{Swain19_NewClass_Terrestrial_Planets}
{Swain}, M.~R., {Estrela}, R., {Sotin}, C., {Roudier}, G.~M., \& {Zellem},
  R.~T. 2019, \apj, 881, 117, \dodoi{10.3847/1538-4357/ab2714}

\bibitem[{{Tsiaras} {et~al.}(2016){Tsiaras}, {Rocchetto}, {Waldmann}, {Venot},
  {Varley}, {Morello}, {Damiano}, {Tinetti}, {Barton}, {Yurchenko}, \&
  {Tennyson}}]{tsiaras16_55Cnce}
{Tsiaras}, A., {Rocchetto}, M., {Waldmann}, I.~P., {et~al.} 2016, \apj, 820,
  99, \dodoi{10.3847/0004-637X/820/2/99}

\bibitem[{{Unterborn} {et~al.}(2018){Unterborn}, {Desch}, {Hinkel}, \&
  {Lorenzo}}]{CTU18_Nature_Trappist}
{Unterborn}, C.~T., {Desch}, S.~J., {Hinkel}, N.~R., \& {Lorenzo}, A. 2018,
  Nature Astronomy, 2, 297, \dodoi{10.1038/s41550-018-0411-6}

\bibitem[{{Unterborn} {et~al.}(2016){Unterborn}, {Dismukes}, \&
  {Panero}}]{CTU16_ApJ_Scaling_the_Earth}
{Unterborn}, C.~T., {Dismukes}, E.~E., \& {Panero}, W.~R. 2016, \apj, 819, 32,
  \dodoi{10.3847/0004-637X/819/1/32}

\bibitem[{Unterborn \& Panero(2017)}]{CTU17oxidation}
Unterborn, C.~T., \& Panero, W.~R. 2017, The Astrophysical Journal, 845, 61

\bibitem[{Unterborn \& Panero(2019)}]{unterborn_panero19}
---. 2019, Journal of Geophysical Research: Planets, 124, 1704,
  \dodoi{10.1029/2018JE005844}

\bibitem[{{Van Eylen} {et~al.}(2018){Van Eylen}, {Agentoft}, {Lundkvist},
  {Kjeldsen}, {Owen}, {Fulton}, {Petigura}, \& {Snellen}}]{VanEylen18}
{Van Eylen}, V., {Agentoft}, C., {Lundkvist}, M.~S., {et~al.} 2018, \mnras,
  479, 4786, \dodoi{10.1093/mnras/sty1783}

\bibitem[{{Vanderburg} {et~al.}(2017){Vanderburg}, {Becker}, {Buchhave},
  {Mortier}, {Lopez}, {Malavolta}, {Haywood}, {Latham}, {Charbonneau},
  {L{\'o}pez-Morales}, {Adams}, {Bonomo}, {Bouchy}, {Collier Cameron},
  {Cosentino}, {Di Fabrizio}, {Dumusque}, {Fiorenzano}, {Harutyunyan},
  {Johnson}, {Lorenzi}, {Lovis}, {Mayor}, {Micela}, {Molinari}, {Pedani},
  {Pepe}, {Piotto}, {Phillips}, {Rice}, {Sasselov}, {S{\'e}gransan},
  {Sozzetti}, {Udry}, \& {Watson}}]{Vanderburg_WASP47e}
{Vanderburg}, A., {Becker}, J.~C., {Buchhave}, L.~A., {et~al.} 2017, \aj, 154,
  237, \dodoi{10.3847/1538-3881/aa918b}

\bibitem[{{Vissapragada} {et~al.}(2020){Vissapragada}, {Jontof-Hutter},
  {Shporer}, {Knutson}, {Liu}, {Thorngren}, {Lee}, {Chachan}, {Mawet},
  {Millar-Blanchaer}, {Nilsson}, {Tinyanont}, {Vasisht}, \&
  {Wright}}]{Vissapragada20_kep36b_MR}
{Vissapragada}, S., {Jontof-Hutter}, D., {Shporer}, A., {et~al.} 2020, \aj,
  159, 108, \dodoi{10.3847/1538-3881/ab65c8}

\bibitem[{{Wang} {et~al.}(2019){Wang}, {Lineweaver}, \&
  {Ireland}}]{Wang19_ElementalAbund_Vol_Trend}
{Wang}, H.~S., {Lineweaver}, C.~H., \& {Ireland}, T.~R. 2019, \icarus, 328,
  287, \dodoi{10.1016/j.icarus.2019.03.018}

\bibitem[{{Wang} \& {Dai}(2019)}]{superPuff}
{Wang}, L., \& {Dai}, F. 2019, \apjl, 873, L1, \dodoi{10.3847/2041-8213/ab0653}

\bibitem[{{Wanke} \& {Dreibus}(1994)}]{Wanke94_Chem_and_Acc_Hist_Mars}
{Wanke}, H., \& {Dreibus}, G. 1994, Philosophical Transactions of the Royal
  Society of London Series A, 349, 285, \dodoi{10.1098/rsta.1994.0132}

\bibitem[{{Winn} {et~al.}(2011){Winn}, {Matthews}, {Dawson}, {Fabrycky},
  {Holman}, {Kallinger}, {Kuschnig}, {Sasselov}, {Dragomir}, {Guenther},
  {Moffat}, {Rowe}, {Rucinski}, \& {Weiss}}]{Winn_55Cnce}
{Winn}, J.~N., {Matthews}, J.~M., {Dawson}, R.~I., {et~al.} 2011, \apjl, 737,
  L18, \dodoi{10.1088/2041-8205/737/1/L18}

\bibitem[{{Yoshizaki} \& {McDonough}(2020)}]{Yoshizaki_McDonough20}
{Yoshizaki}, T., \& {McDonough}, W.~F. 2020, \gca, 273, 137,
  \dodoi{10.1016/j.gca.2020.01.011}

\bibitem[{{Zharkov}(1983)}]{Zharkov83_IntStruct_Venus}
{Zharkov}, V.~N. 1983, Moon and Planets, 29, 139, \dodoi{10.1007/BF00928322}

\bibitem[{{Zharkov} \& {Gudkova}(2005)}]{Zharkov_Gudkova_2005}
{Zharkov}, V.~N., \& {Gudkova}, T.~V. 2005, Solar System Research, 39, 343,
  \dodoi{10.1007/s11208-005-0049-7}

\end{thebibliography}
\bibliographystyle{aasjournal}

\section*{Supplementary Information and Supporting Data}

\counterwithin{figure}{section}

\setcounter{table}{0}
\setcounter{page}{1}
\setcounter{figure}{0}

\renewcommand{\thepage}{S\arabic{page}} 
\renewcommand{\thesection}{S\arabic{section}}  
\renewcommand{\thetable}{S\arabic{table}}  
\renewcommand{\thefigure}{S\arabic{figure}}

\subsection*{Calculation of CMF and $\sigma_{\text{CMF}}$}

\noindent  CMF$_\rho$, CMF$_{\mathrm{star}}$, and their uncertainties are calculated based on planetary mass, radius, host\hyp{}star Fe/Mg and Si/Mg as well as the uncertainty in each parameters. Example models in Table \ref{dcmfrhotab} and \ref{dcmfstar} were calculated using \texttt{ExoLens}, an open\hyp{}source compositional calculator for rocky planets based on \texttt{ExoPlex}. This calculator estimates the core mass fractions, CMF$_\rho$, CMF$_{\mathrm{star}}$, and their uncertainties of a 0.1\hyp{}10$M_\oplus$ planet from mass and radius to within 1\hyp{}2\% of \texttt{ExoPlex}, automatically calculating $P(\mathcal{H}^0)$. \texttt{ExoLens} is available at the GitHub (\href{https://github.com/schulze61/ExoLens}{https://github.com/schulze61/ExoLens}).

\begin{table}[ht]
    
    \centering
    \begin{tabular}{|c|c|c|c|c|c|c|}
    \hline
     M$_p$ & $\sigma_{M_p}$ ($\%$) & R$_p$ & $\sigma_{R_p}$ ($\%$) & CMF$_\rho$ & $\sigma_{\text{CMF}_\rho, upper}$ & $\sigma_{\text{CMF}_\rho, lower}$  \\

    \hline
    5 & 10 & 1.54 & 1 & 0.35 & 0.09 & 0.10  \\
    5 & 10 & 1.54 & 2.5 &  0.35 & 0.12 & 0.13  \\
    5 & 10 & 1.54 & 10 & 0.35 & 0.32 & 0.39  \\
    
     \hline
    5 & 5 & 1.54 & 5 & 0.35 & 0.17 & 0.19 \\
    5 & 10 & 1.54 & 5 & 0.35 & 0.18 & 0.20 \\
    5 & 20 & 1.54 & 5 & 0.35 & 0.22 & 0.28 \\
    
    \hline
    1 & 10 & 0.99 & 5 & 0.35 & 0.19 & 0.22  \\
    2.5 & 10 & 1.28 & 5 & 0.35 & 0.19 & 0.21 \\
    7.5 & 10 & 1.71 & 5 & 0.35 & 0.18 & 0.20 \\
    10 & 10 & 1.84 & 5 & 0.35 & 0.17 & 0.20 \\
    
    \hline
    5 & 10 & 1.65 & 5 & 0.10 & 0.22 & $>$0.10  \\
    5 & 10 & 1.59 & 5 & 0.25 & 0.20 & 0.22 \\
    5 & 10 & 1.50 & 5 & 0.45 & 0.17 & 0.19  \\
    5 & 10 & 1.37 & 5 & 0.70 & 0.13 & 0.14\\
    5 & 10 & 1.25 & 5 & 0.90 & 0.10 & 0.11  \\

     \hline
    \end{tabular}
    \caption{The effects of $M_p$, $R_p$, $\sigma_{R_p}$ and $\sigma_{M_p}$ on CMF$_\rho$ and $\sigma_{\text{CMF}_\rho}$.}
    
    \label{dcmfrhotab}
\end{table}

\begin{table}[ht]
    \centering
        \begin{tabular}{|c|c|c|c|c|c|c|}
            \hline
            Fe/Mg & $\sigma_{\mathrm{Fe/Mg}}$ ($\%$) & Si/Mg & $\sigma_{\mathrm{Si/Mg}}$ ($\%$) & CMF$_{\mathrm{star}}$ & $\sigma_{\text{CMF}_{\mathrm{star}}}$ \\
    
            \hline
            1.0 & 20 & 1.0 & 10 & 0.36 & 0.05 \\
            1.0 & 20 & 1.0 & 20 & 0.36 & 0.05 \\
            1.0 & 20 & 1.0 & 30 & 0.36 & 0.06 \\
            1.0 & 20 & 1.0 & 40 & 0.36 & 0.07 \\
            1.0 & 20 & 1.0 & 50 & 0.36 & 0.09 \\
    
            \hline
            1.0 & 20 & 0.5 & 20 & 0.44 & 0.05 \\
            1.0 & 20 & 0.8 & 20 & 0.39 & 0.05 \\
            1.0 & 20 & 1.25 & 20 & 0.33 & 0.05 \\
            1.0 & 20 & 2.0 & 20 & 0.26 & 0.05 \\

            \hline
            1.0 & 10 & 1.0 & 20 & 0.36 & 0.04 \\
            1.0 & 30 & 1.0 & 20 & 0.36 & 0.07 \\
            1.0 & 40 & 1.0 & 20 & 0.36 & 0.10 \\
            1.0 & 50 & 1.0 & 20 & 0.36 & 0.12 \\
    
            \hline
            0.5 & 20 & 1.0 & 20 & 0.22 & 0.04\\
            0.8 & 20 & 1.0 & 20 & 0.31 & 0.07  \\
            1.25 & 20 & 1.0 & 20 & 0.41 & 0.10 \\
            2.0 & 20 & 1.0 & 20 & 0.53 & 0.12 \\
     
            \hline
        \end{tabular}
    
    \caption{The effects of Fe/Mg, Si/Mg, $\sigma_{\mathrm{Fe/Mg}}$ and $\sigma_{\mathrm{Si/Mg}}$ on $\mathrm{CMF_\star}$ and $\sigma_{\text{CMF}_{\mathrm{star}}}$.}
    \label{dcmfstar}
\end{table}

\newpage
\subsection*{Previous studies}
\noindent We calculate CMF$_\rho$ estimates for planets in our sample with multiple reported mass and radius measurements on \href{https://exoplanetarchive.ipac.caltech.edu/}{NASA Exoplanet Archive}. We only consider studies where both mass and radius, and thus bulk density, are measured. All CMF$_\rho$ values presented here are calculated using \texttt{ExoLens}.

\begin{table}[ht]

    \centering
    \begin{tabular}{|c|c|c|c|c|c|}
    \hline
         Planet &  R$_p$ [$R_\oplus$] & M$_p$ [$M_\oplus$]  & Source & CMF$_\rho$ & $P(\mathcal{H}^0)$ (\%)\\
         \hline
         K2\hyp{}229 b &  $1.164^{+0.066}_{-0.048}$ & $2.59\pm 0.43$ & \citet{Santerne18_K2229b} & $0.68^{+0.15}_{-0.18}$ & 13\\
         \hline
         HD 219134 c & $1.511\pm 0.047$ & $4.36 \pm 0.22$ &  \citet{HD213194bc_Gillon} & $0.30^{+0.12}_{-0.13}$ & 99\\
         \hline
         Kepler\hyp{}10 b &  $1.481^{+0.049}_{-0.029}$ & $4.61^{+1.27}_{-1.26}$ & \citet{Esteves_Kep10b} & $0.42^{+0.21}_{-0.35}$ & 90 \\
         & $1.47\pm0.03$ & $3.33 \pm 0.49$ &  \citet{Dumusque_Kep10b} & $0.12^{+0.17}_{<0}$ & 74\\
         &  $1.46\pm0.02$ & $4.60\pm1.26$ & \citet{Fogtmann_kep10b} & $0.46^{+0.18}_{-0.30}$ & 81\\
         & $1.416^{+0.033}_{-0.036}$ & $4.56^{+1.17}_{-1.29}$ & \citet{Batalha_kep10b} & $0.55^{+0.17}_{-0.27}$ & 58\\
         \hline
         HD 219134 b & $1.602\pm0.055$ & $4.74\pm0.19$ & \citet{HD213194bc_Gillon} & $0.15^{+0.14}_{-0.15}$ & 71\\
         & $1.606\pm 0.086$ & $4.36 \pm 0.44$ & \citet{HD219134b_Motalebi_MR} & $0.05^{+0.24}_{<0}$ & 65 \\
         \hline
         HD 15337 b & $1.64\pm0.06$ & $7.51^{+1.09}_{-1.01}$ & \citet{Gandolfi19_HD15337b} & $0.49^{+0.14}_{-0.16}$ & 48\\
         \hline
         WASP\hyp{}47 e &  $1.810\pm0.027$ & $6.83\pm0.66$ & \citet{Vanderburg_WASP47e} & $0.04^{+0.12}_{<0}$ & 24\\
         &  $1.82\pm0.40$ &  $9.10^{+5.50}_{-3.60}$ & \citet{Almenara_WASP47e} & $0.31^{+0.66}_{<0}$ & 100\\
         &  $1.817\pm0.065$ & $12.2\pm 3.7$ & \citet{Dai15_WASP47e} & $0.55^{+0.18}_{-0.29}$ & 62 \\
         \hline
         Kepler\hyp{}20 b &  $1.91^{+0.12}_{-0.21}$ &  $8.7^{+2.1}_{-2.2}$ & \citet{Gautier12_Kep20b} & $0.08^{+0.39}_{<0}$ &  84 \\
         \hline 
         55 Cnc e &  $1.947\pm0.038$ & $8.59\pm0.43$ & \citet{Crida_2018} & $[<0, 0.08]$ & 4\\
         & $1.875\pm0.029$ & $7.99^{+0.32}_{-0.33}$ & \citet{Bourrier18_55Cnce} & $0.06^{+0.08}_{-0.08}$ & 14\\
         & $1.91 \pm 0.08$ & $8.08\pm0.31$ & \citet{Demory16_55Cnce} & $0^{+0.18}$ & 31\\
         & $2.173^{+0.097}_{-0.098}$ & $8.37\pm0.38$ & \citet{Endl_55Cnce} & $<0$ & 0\\
         & $2.08^{+0.16}_{-0.17}$ & $7.81^{+0.58}_{-0.53}$ & \citet{Demory11_55Cnce} & $<0$ & 0 \\
         & $2.00\pm0.14$ & $8.63\pm0.35$ &  \citet{Winn_55Cnce} & $[<0, 0.17]$ & 40\\
         \hline

    \end{tabular}
    \caption{CMF$_\rho$ values for previously reported mass and radius measurements in order of publication date.}
    \label{tab:prev_studies}
\end{table}

\newpage
\subsection*{Minimum Uncertainties}
\begin{table}[ht]

    \centering
    \begin{tabular}{|c|c|c|c|c|c|c|c|}
        \hline
         Planet & CMF$_\rho$ & CMF$_{\mathrm{star}}$ & $\sigma_{\mathrm{Fe/Mg}} = \sigma_{\mathrm{Si/Mg}}$ (\%) & $P(\mathcal{H}^0)$ (\%) & 1$\sigma$ Class & 2$\sigma$ Class\\
         \hline
         K2\hyp{}229 b &  $0.565^{+0.041}_{-0.043}$ & $0.29\pm0.06$ & -- & 0.1 & SM & SM\\
         HD 219134 c &  $0.42^{+0.05}_{-0.05}$ &  $0.28 \pm 0.03$ & 14 & 5 & SM & SM\\
         Kepler\hyp{}10 b &  $0.13^{+0.06}_{-0.06}$ & $0.28 \pm 0.02$ & 9 & 5 & LDSP & LDSP\\ 
         HD 219134 b &  $0.29 ^{+0.05}_{-0.05}$ & $0.28 \pm 0.02$ & 8 & 98 & IHS & IHS\\
         Kepler\hyp{}107 c &  $0.70^{+0.03}_{-0.03}$ & $0.30\pm 0.07$ & -- & $\sim 0$ & SM & SM\\
         HD 15337 b &  $0.34^{+0.05}_{-0.05}$ & $0.29\pm 0.02$ & 8 & 69 & IHS & IHS \\
         K2\hyp{}265 b & $0.24^{+0.04}_{-0.05}$ & $0.33\pm 0.02$ & 8 & 13 & LDSP & IHS\\
         HD 213885 b & $0.42^{+0.04}_{-0.05}$ & $0.31\pm 0.02$ & 8 & 12 & SM & IHS\\        
         WASP\hyp{}47 e &  $0.155^{+0.056}_{-0.058}$ & $0.26 \pm 0.02$ & 8 & 21 & LDSP & IHS\\
         Kepler\hyp{}20 b &  $0.26^{+0.05}_{-0.05}$ & $0.30\pm 0.01$ & 8 & 76 & IHS & IHS\\
         55 Cnc e &  $0.004^{+0.06}_{<0}$ & $0.31\pm 0.10$ & -- & 3.4 & LDSP & LDSP\\
         \hline
    \end{tabular}
    \caption{Minimum uncertainties needed to classify the planets in our sample at the 2$\sigma$ significance level. For these calculations, we assume $\sigma_{M_p} = 4\%$ and $\sigma_{R_p} = 1\%$. We vary $\sigma_{\mathrm{Fe/Mg}} = \sigma_{\mathrm{Si/Mg}}$ until $P(\mathcal{H}^0) \leq 5\%$ stopping at the optimistic value of $\sigma_{\mathrm{Fe/Mg}} = \sigma_{\mathrm{Si/Mg}} = 8\%$ if this $P(\mathcal{H}^0)$ criteria is not met. For these planets, the $P(\mathcal{H}^0)$ corresponding to $\sigma_{M_p} = 4\%$, $\sigma_{R_p} = 1\%$, and $\sigma_{\mathrm{Fe/Mg}} = \sigma_{\mathrm{Si/Mg}} =  8\%$ is recorded. This stellar abundance criteria is derived from the recommended abundances of \citet{Lodders03}. While such precisions will be difficult to obtain for extrasolar stars from spectroscopic abundance measurements alone, $\sigma_{\mathrm{Fe/Mg}} = \sigma_{\mathrm{Si/Mg}} =  8\%$, nonetheless, provides a good stopping point for our minimum precision calculations. Planets with no value recorded in the abundance ratio uncertainty column do \textit{not} require increased abundance precisions to classify at better than the 2$\sigma$ level if their mass and radius uncertainties reach 4\% and 1\%, respectively. }
    \label{tab:min_unc_table}
\end{table}

\newpage
\begin{sidewaysfigure}
\subsection*{Calculating $P(\mathcal{H}^0)$: Schematic for quantitative rejection of the null hypothesis}

    \begin{tabular}{c}
        \includegraphics[width=0.9\textwidth]{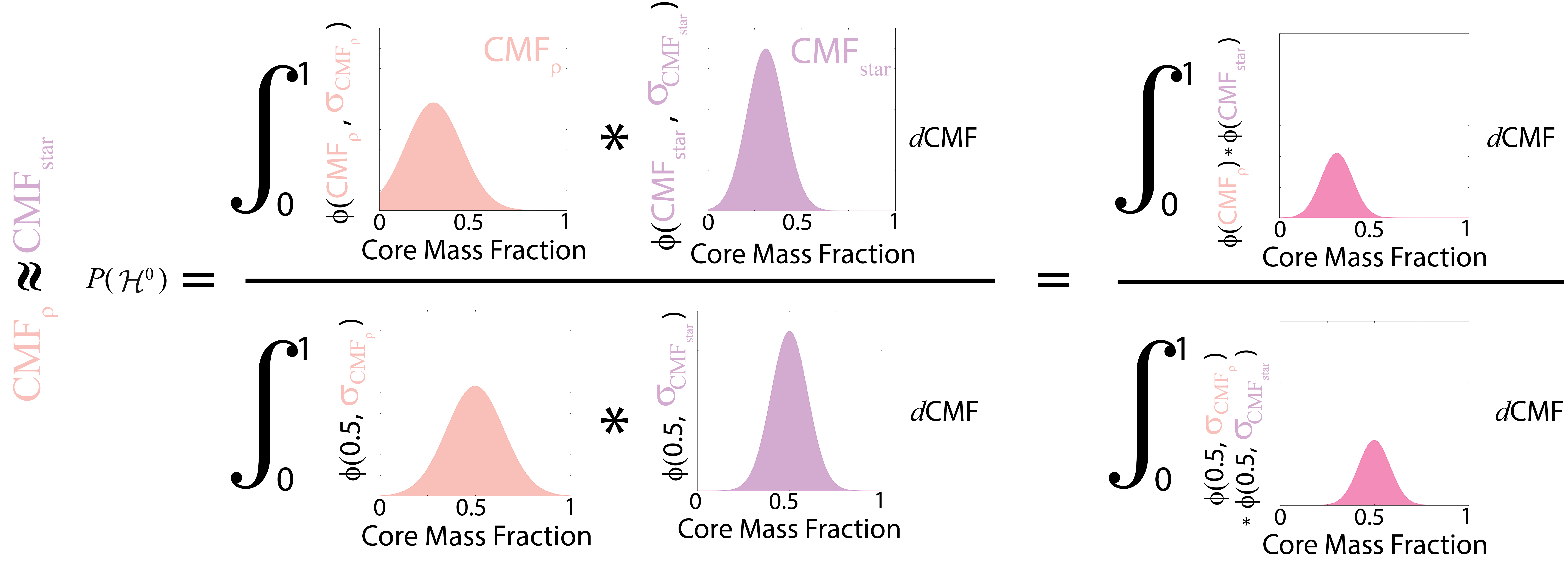}\\
        \includegraphics[width=0.9\textwidth]{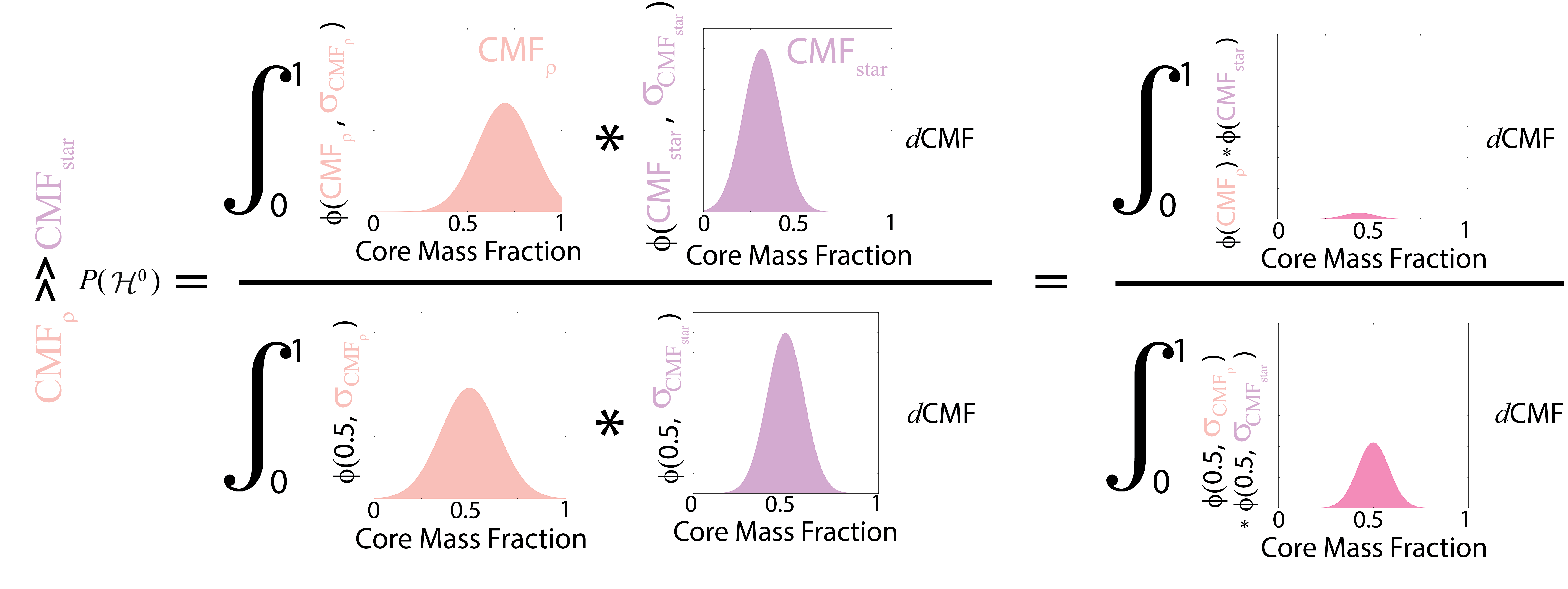}\\
    \end{tabular}
    \caption{We illustrate the meaning of Equation (\ref{equ:probability}) with an example of a planet whose CMF$_\rho$ is similar to its CMF$_{\mathrm{star}}$ value (top) and one whose CMF$_\rho$ is greater than its CMF$_{\mathrm{star}}$ value (bottom). The numerator of eqn. \ref{equ:probability} computes the integral of the product of the distribution of each CMF value. In the case where they are similar, the overlaps are significant, yet where the CMF values are different, there is nearly no overlap. The denominator for each case accounts for the consequences of variable distribution widths so as to normalize the numerator to 1.}
    \label{fig:eq3_schematic}

\end{sidewaysfigure}

\newpage
\subsection*{Figure 4 Planet Sample}

\begin{table}[ht]
   \centering
    
    \tiny
    \begin{tabular}{|c|c|c|c|c|c|c|c|c|}
    \hline

     Planet & R$_p$ [$R_\oplus$] & M$_p$ [$M_\oplus$] & M\hyp{}R Source & P [days] & S.T. & Fe/Mg & Si/Mg & Spect. Source  \\
     \hline
     K2\hyp{}229 b & $1.197^{+0.045}_{-0.048}$ (3.9\%) & $2.49^{+0.42}_{-0.43}$ (17.1\%) & \citet{Dai_Homogenous_MRs_HotEarths} & 0.58 & G9 & 0.78$\pm$0.19 & 1.1$\pm$0.24 & \citet{Santerne18_K2229b}\\
     HD 219134 c & 1.415$\pm$0.049 (3.5\%) & 3.96$\pm$0.34 (8.6\%) &  \citet{Ligi19_HD219134} & 6.76 & K3 & 0.69$\pm$0.25 & 0.98$\pm$ 0.39 & Hypatia Catalog \\
     Kepler\hyp{}10 b & $1.489^{+0.023}_{-0.021}$ (1.5\%)  & $3.57^{+0.51}_{-0.53}$ (14.6\%) & \citet{Dai_Homogenous_MRs_HotEarths} & 0.84 & G & 0.62$\pm$0.14 &0.83$\pm$0.16  & \citet{Liu16_kep10_abunds}\\
     HD 219134 b & 1.500$\pm$0.057 (3.8\%) & 4.27$\pm$0.34 (8.0\%)  &  \citet{Ligi19_HD219134} & 3.09 & K3 & 0.69$\pm$0.25 & 0.98$\pm$ 0.39 & Hypatia Catalog\\
     Kepler\hyp{}107 c  & 1.597$\pm$0.026 (18.9\%) &  9.39$\pm$1.77 (1.6\%) & \citet{Bonomo19_Kep107_Nature} &  4.9 & G2 & 0.75$\pm$ 0.22 & 0.96$\pm$0.23 & \citet{Bonomo19_Kep107_Nature}\\
     HD 15337 b & $1.699^{+0.062}_{-0.059}$ (3.6\%) & $7.20\pm 0.81$ (11.3\%)  & \citet{HD15337_MR_Dumusque} & 4.76 & K1 & 0.69$\pm$0.29 & 0.87$\pm$ 0.20 & Hypatia Catalog\\ 
     K2\hyp{}265 b & $1.71 \pm 0.11$ (6.4\%) & $6.54 \pm 0.84$ (12.8\%) & \citet{Lam18_k2265b} & 2.37 & G8 & 0.84$\pm0.24$ & 0.92$\pm0.24$ & \citet{Lam18_k2265b} \\
     HD 213885 b  & $1.745^{+0.051}_{-0.052}$ (3.0\%)  & $8.83^{+0.66}_{-0.65}$ (7.4\%)   & \citet{Espinoza_HD213885} & 1.008 & G & 0.81$\pm$0.23 & 0.98$\pm$0.31 & \citet{Espinoza_HD213885} \\ 
     WASP\hyp{}47 e & $1.773^{+0.049}_{-0.048}$ (2.7\%)  & $6.91^{+0.81}_{-0.83}$ (11.9\%)  & \citet{Dai_Homogenous_MRs_HotEarths} &  0.79 & G9 & 0.76$\pm$0.22 & 1.35$\pm$0.36 & \citet{Hellier12_WASP47} \\
     Kepler\hyp{}20 b & $1.868^{+0.066}_{-0.034}$ (2.7\%) & $9.70^{+1.41}_{-1.44}$ (14.7\%)  & \citet{Buchhave16_Kep20b_MR} &  3.70 & G8 & 0.71$\pm$0.27 & 0.90$\pm$0.41 & \citet{Kep20_abundances} \\
     55 Cnc e  & $1.897^{+0.044}_{-0.046}$ (2.4\%) & $7.74^{+0.37}_{-0.30}$ (4.3\%)   & \citet{Dai_Homogenous_MRs_HotEarths} & 0.74 & G8 &  0.76$\pm$ 0.32 & 0.87$\pm$0.34 & Hypatia Catalog \\
     \hline
     GJ 1132 b & $1.130\pm0.056$ (5.0\%) & $1.66\pm0.23$ (13.9\%) & \citet{Bonfils18_GJ1132MR} & 1.63 & M4.5 & -- & -- & --\\
     GJ 357 b & $1.217^{+0.084}_{-0.083}$ (6.9\%)  & $1.84\pm0.31$ (16.9\%) & \citet{Luque18_GJ357b_MR} & 3.93 & M2.5 & -- & -- & --\\
     LTT 3780 b & $1.332^{+0.072}_{-0.075}$ (5.5\%)& $2.62^{+0.48}_{-0.46}$ (17.9\%) & \citet{Cloutier20_LT3780_MR} & 0.77 & M4 & -- & -- & --\\
     Kepler-105 c & $1.31 \pm 0.07$ (5.3\%) & $4.60^{+0.92}_{-0.85}$ (19.2\%) & \citet{JontofHutter16_Kep60b__kep105c_MR} & 7.13 &G1 & -- & -- & --\\
     L 98-59 c & $1.35\pm0.07$ (5.2\%) & $2.42^{+0.35}_{-0.34}$ (14.3\%) & \citet{Cloutier19_L9858cd_MR} & 3.69 & M3 & -- & -- & --\\
     L 168-9 b & $1.39\pm0.09$ (6.5\%) & $4.60\pm0.56$ (12.2\%) & \citet{AD20_L1689b_MR} & 1.40 & M1 & -- & -- & --\\
     Kepler-406 b & $1.43\pm 0.03$ (2.1\%) & $6.35\pm1.4$ (22.1\%) & \citet{Marcy14_kep406b_MR} & 2.43 & G7 & -- & -- & --\\
     Kepler-36 b & $1.498^{+0.061}_{-0.049}$ (3.7\%) & $3.83^{+0.11}_{-0.10}$ (2.7\%) & \citet{Vissapragada20_kep36b_MR} & 13.87 & G1 & -- & -- & --\\
     K2-141 b &$1.51\pm 0.05$ (3.3\%) & $5.08\pm0.41$ (8.1\%) & \citet{Malavolta18_k2141bMR} & 0.28 & K7 & -- & -- & --\\
     Kepler-80 d & $1.53^{+0.09}_{-0.07}$ (5.2\%) & $6.75^{+0.69}_{-0.51}$ (8.9\%) & \citet{MacDonald16_Kep80d_MR} & 3.07 & K5 & -- & -- & --\\
     L 98-59 d & $1.57\pm0.14$ (8.9\%) & $2.31^{+0.46}_{-0.45}$ (19.7\%) & \citet{Cloutier19_L9858cd_MR} & 7.45 & M3 & -- & -- & --\\
     GJ 9827 b & $1.577^{+0.027}_{-0.031}$ (1.8\%) & $4.91\pm0.49$ (10.0\%) & \citet{Rice19_GJ9827b_MR} & 1.21 & K5 & -- & -- & --\\
     K2-291 b & $1.589^{+0.095}_{-0.072}$ (5.3\%) & $6.49\pm1.16$ (17.9\%) & \citet{Kosiarek19_K2291b_MR} & 2.23 & G7 & -- & -- & --\\
     HD 80653 b & $1.613\pm0.071$ (4.4\%) & $5.60\pm0.43$ (7.7\%) & \citet{Frustagli20_HD80653b_MR} & 0.72 & K5 & -- & -- & --\\
     Kepler-60 b & $1.71\pm0.13$ (7.6\%) & $4.19^{+0.56}_{-0.52}$ (12.9\%) & \citet{JontofHutter16_Kep60b__kep105c_MR} & 7.13 & G1 & -- & -- & --\\
     TOI-1235 b & $1.738^{+0.087}_{-0.076}$ (4.7\%) & $6.91^{+0.75}_{-0.85}$ (11.6\%) & \citet{Cloutier20_TOI1235b_MR} & 3.45 & M0.5 & -- & -- & --\\
     K2-216 b & $1.75^{+0.17}_{-0.10}$ (7.7\%) & $8.0\pm1.6$ (20\%) & \citet{Persson18_K2216b_MR} & 2.17 & K5 & -- & -- & -- \\

     \hline
    \end{tabular}
    
    \caption{Sample of well\hyp{}characterized exoplanets including those without reported host star abundance measurements beyond [Fe/H] that otherwise meet our selection criteria. Host star elemental ratios Fe/Mg and Si/Mg are expressed as molar ratios. For each star, we derive molar ratios of Fe/Mg and Si/Mg using the solar abundances from \citet{Lodders09}. Planets with known host-star Fe/Mg and Si/Mg values are separated from those with unknown host Fe/Mg and Si/Mg via a horizontal line. Planets are listed in order of increasing radius within each of these subsets. S.T. = Spectral Type of the host star.}
    
    \label{table:AllpSAMPLE}
\end{table}

\end{CJK*}

\listofchanges
\end{document}